\documentclass[fleqn,usenatbib]{mnras}

\usepackage{newtxtext,newtxmath}

\usepackage[T1]{fontenc}
\usepackage[dvipsnames]{xcolor}
\usepackage{booktabs}
\usepackage{lscape}
\DeclareRobustCommand{\VAN}[3]{#2}
\let\VANthebibliography\thebibliography
\def\thebibliography{\DeclareRobustCommand{\VAN}[3]{##3}\VANthebibliography}

\usepackage{graphicx}	
\usepackage{amsmath}	

\title[Circular velocity and halo mass functions]{The circular velocity and halo mass functions of galaxies in the nearby Universe}

\author[A. Ristea et al.]{
Andrei Ristea,$^{1,2}$\thanks{E-mail: andrei.ristea@research.uwa.edu.au (UWA)}
Luca Cortese,$^{1,2}$
Brent Groves,$^{1,2}$
A. Fraser-McKelvie,$^{2,3}$
Danail Obreschkow$^{1,2}$  
\newauthor
\ and Karl Glazebrook$^{4}$
\\
$^{1}$International Centre for Radio Astronomy Research, The University of Western Australia, 35 Stirling Highway, Crawley WA 6009, Australia\\
$^{2}$ARC Centre of Excellence for All Sky Astrophysics in 3 Dimensions (ASTRO 3D), Australia\\
$^{3}$ European Southern Observatory, Karl-Schwarzschild-Straße 2, 85748 Garching, Germany \\
$^{4}$Centre for Astrophysics and Supercomputing, Swinburne University of Technology, Hawthorn, VIC 3122, Australia
}

\date{Accepted XXX. Received YYY; in original form ZZZ}

\pubyear{2015}

\begin{document}
\label{firstpage}
\pagerange{\pageref{firstpage}--\pageref{lastpage}}
\maketitle

\begin{abstract}
The circular velocity function (CVF) of galaxies is a fundamental test of the $\Lambda$ Cold Dark Matter (CDM) paradigm as it traces the variation of galaxy number densities with circular velocity ($v_{\rm{circ}}$), a proxy for dynamical mass. Previous observational studies of the CVF have either been based on \ion{H}{I}-rich galaxies, or encompassed low-number statistics and probed narrow ranges in $v_{\rm{circ}}$. We present a benchmark computation of the CVF between $100-350\ \rm{km\ s^{-1}}$ using a sample of 3527 nearby-Universe galaxies, representative for stellar masses between $10^{9.2}-10^{11.9} \rm{M_{\odot}}$. We find significantly larger number densities above 150 $\rm{km\ s^{-1}}$ compared to results from \ion{H}{I} surveys, pertaining to the morphological diversity of our sample.
Leveraging the fact that circular velocities are tracing the gravitational potential of halos, we compute the halo mass function (HMF), covering $\sim$1 dex of previously unprobed halo masses ($10^{11.7}-10^{12.7} \rm{M_{\odot}}$). The HMF for our sample, representative of the galaxy population with $M_{200}\geqslant10^{11.35} \rm{M_{\odot}}$, shows that spiral morphologies contribute 67 per cent of the matter density in the nearby Universe, while early types account for the rest.
We combine our HMF data with literature measurements based on \ion{H}{I} kinematics and group/cluster velocity dispersions. We constrain the functional form of the HMF between $10^{10.5}-10^{15.5} \rm{M_{\odot}}$, finding a good agreement with $\Lambda$CDM predictions. The halo mass range probed encompasses 72$\substack{+5 \\ -6}$ per cent ($\Omega_{\rm{M,10.5-15.5}} = 0.227 \pm 0.018$) of the matter density in the nearby Universe; 31$\substack{+5 \\ -6}$ per cent is accounted for by halos below $10^{12.7}\rm{M_{\odot}}$ occupied by a single galaxy.

\end{abstract}




\section{Introduction}

Galactic evolutionary processes acting across cosmic time produce a galaxy population in the nearby Universe with certain statistical properties. 
Reproducing galaxy population statistics such as the stellar mass and luminosity functions (i.e., the number density of galaxies as a function of stellar mass and luminosity) in the present-day Universe is a paramount test of the $\Lambda$CDM cosmological framework.

The stellar mass function has been the object of intense study since the advent of spectroscopic surveys (see e.g. \citealt{binggeli_luminosity_1988}, \citealt{gardner_wide-field_1997} , \citealt{baldry_galaxy_2012}, \citealt{moffett_galaxy_2016}, \citealt{wright_galaxy_2017}, \citealt{thorne_deep_2021}). Its shape and amplitude have been well constrained over several orders of magnitude, and the contribution of different morpholologies to the number density of galaxies at different stellar masses has been comprehensively charted (\citealt{guo_sami_2020}, \citealt{driver_galaxy_2022}). Reproducing the stellar mass function has been the benchmark of numerical simulations of galaxy evolution based on $\Lambda$CDM theory (see \citealt{wechsler_connection_2018} for a review). However, the stellar mass budget of galaxies is influenced by baryonic processes (e.g. gas accretion, feedback, star formation efficiency) with technicalities that have yet to be fully understood and modelled theoretically.

A galaxy property related to the halo mass of the system, which is less affected by baryonic physical processes than stellar mass, is the circular velocity $v_{\rm{circ}}$.
Several definitions of $v_{\rm{circ}}$ are used in numerical galaxy simulations, most notably the maximum circular velocity of halos (\citealt{kravtsov_dark_2004}, \citealt{zavala_velocity_2009}), the velocity at the virial radius and the velocity at 80 per cent of the mass radius (\citealt{bekeraite_califa_2016}). There is, however, no widely accepted method for computing circular velocities in observational studies. Early computations of the circular velocity function (CVF) have used the galactic luminosity function in conjunction with the Tully-Fisher relation (\citealt{tully_new_1977}) and/or the Faber-Jackson relation (\citealt{faber_velocity_1976}) to estimate circular velocities (\citealt{gonzalez_velocity_2000}, \citealt{desai_cluster_2004}, \citealt{abramson_circular_2014}). A similar approach has been taken when computing the related velocity dispersion function for early-type galaxies (\citealt{sheth_velocity_2003}, \citealt{choi_internal_2007}, \citealt{chae_galaxy_2010}). However, the CVFs and velocity dispersion functions computed in this manner are very sensitive to the assumptions regarding luminosity functions and luminosity-velocity scaling relations.

More recently, CVF studies have relied on direct kinematic measurements from two blind \ion{H}{I} surveys: The \ion{H}{I} Parkes All Sky (HIPASS) survey and the Arecibo Legacy Fast ALFA (ALFALFA) survey. Using these surveys, \cite{zwaan_velocity_2010} (HIPASS) and \cite{papastergis_velocity_2011} (ALFALFA) have constrained the CVF for gas-rich galaxies in the range $50-350\ \rm{km\ s^{-1}}$. By combining \ion{H}{I} data with photometric velocity estimators for a local-volume sample, \cite{klypin_abundance_2015} have extended the CVF down to $15\ \rm{km\ s^{-1}}$ (albeit only reaching a maximum velocity of $200\ \rm{km\ s^{-1}}$). CVF computations based on \ion{H}{I} surveys however come with the caveat that they only represent gas-rich galaxies and do not include gas-poor early-types. \cite{bekeraite_califa_2016} have computed the CVF in the nearby Universe using a sample which included elliptical morphologies, drawn from the Calar Alto Integral-Field Spectroscopic (CALIFA, \citealt{sanchez_califa_2012}) survey. While morphologically varied, their sample features only 226 galaxies with sizeable velocity uncertainties (as discussed in \citealt{bekeraite_space_2016} and \citealt{bekeraite_califa_2016}), and only probes the range between 160-320 $\rm{km\ s^{-1}}$ with sufficiently high completeness. 

Large integral-field spectroscopic (IFS) surveys in the nearby Universe, most notably the Sloan Digital Sky Survey IV - Mapping nearby Galaxies at Apache Point Observatory (SDSS IV - MaNGA, \citealt{bundy_overview_2015, drory_manga_2015}) and the Sydney-AAO Multi-object Integral-field spectrograph (SAMI) Galaxy Survey \citep{croom_sydney-aao_2012, bryant_sami_2015} offer valuable advantages in terms of computing the CVF. Such surveys measure both the stellar and ionised gas kinematics for morphologically diverse samples of  galaxies, and thus sample both the quenched and active populations. The disadvantage of these IFS surveys is the relatively low extent of the kinematics (1.5 and 2.5 half-light radii for 80 per cent of the Primary and Secondary MaNGA survey samples, respectively, \citealt{bundy_overview_2015}), compared to \ion{H}{I} surveys which probe significantly larger radii (see e.g., \citealt{catinella_sami-h_2023}).  

A natural extension of the CVF is the halo mass function (HMF), under the assumption that circular velocities can be used as proxies for the total halo mass. The functional form of the HMF is one of the most fundamental predictions of the $\Lambda$CDM cosmological model (\citealt{frenk_formation_1988}, \citealt{brainerd_mass_1992}). Numerical simulations of galaxy evolution produce a HMF that can be well described by a  power law (independent of time and scale) up to a cut-off mass (\citealt{jenkins_mass_2001}), followed by an exponential decline. However, the HMF shows significant variations (up to 20 per cent) between different numerical simulations of galaxy evolution (\citealt{murray_how_2013}).

It has been shown by \cite{murray_empirical_2018} that the HMF can be modelled using a four-parameter function, the Murray-Robotham-Power (MRP) function, with an accuracy within 5 per cent at all masses. The MRP is similar to a Schechter (\citealt{schechter_analytic_1976}) function, with the addition of an extra parameter controlling the sharpness of the exponential cut-off in the high mass end. The exponential cut-off mass and sharpness of the HMF are strong probes of the hierarchical assembly process of galaxies: as increasingly larger halos form through merging, the  cut-off shifts to larger masses representative of group and cluster halos (\citealt{allen_cosmological_2011}).
The low mass end of the HMF is strongly dependent on the dark matter particle mass (\citealt{murray_how_2013}). In a $\Lambda$CDM framework, the low mass cut-off of the HMF is around $5\times10^{-6} \rm{M_{\odot}}$ (close to Earth mass), while for warm and hot dark matter, this cut-off can be as high as $10^{13} \rm{M_{\odot}}$ (\citealt{murray_how_2013}). 

Another predictive aspect of the HMF is its integral over all halo masses which provides an estimate of the total matter density. While for cold dark matter all particles are accounted for by halos down to the cut-off mass, in the case of warm or hot dark matter free streaming occurs. Due to this process, a percentage of the matter density predicted by cosmic microwave background studies (CMB, \citealt{aghanim_planck_2020}) will not be bound to dark matter halos.

Observationally, the main avenue for computing the HMF involves identifying group catalogues and estimating their mass from velocity dispersions while assuming that halos are fully virialised. This approach depends greatly on group and cluster identification, which can be achieved from X-ray detections (\citealt{bohringer_extended_2017}) or via group finding algorithms applied to large-scale spectroscopic surveys (e.g. \citealt{driver_empirical_2022}). Both of these approaches however rely on a number of assumptions, most notably the X-ray luminosity - mass relation and the calibration of group-finding algorithms to numerical simulations (see \citealt{driver_empirical_2022} for a comprehensive discussion). 
The HMF for groups and clusters was constrained to a lower halo mass limit of $10^{13.75}\rm{M_{\odot}}$ by \cite{bohringer_extended_2017} via X-ray cluster detections. More recently, \cite{driver_empirical_2022} pushed HMF measurements down to $10^{12.7} \rm{M_{\odot}}$ by combining group and cluster halo mass measurements from the Galaxy and Mass Assembly (GAMA, \citealt{driver_galaxy_2011}) survey, Sloan Digital Sky Survey (SDSS, \citealt{york_sloan_2000}) data release 12 (\citealt{alam_eleventh_2015}, with the final group catalogue being compiled by \citealt{tempel_merging_2017}; see also \citealt{tempel_flux-_2014}) and the ROSAT-ESO Flux Limited X-ray galaxy cluster survey (REFLEX II, \citealt{bohringer_extended_2017}), and resolved $41 \pm 5$ per cent of the matter density in the nearby Universe, with no evidence of a low mass cut-off up to their halo mass limit.

In this work, we provide a benchmark computation of the CVF in the nearby Universe (between $100-350 \ \rm{km\ s^{-1}}$) by combining stellar and gas kinematics for a sample of 3527 galaxies drawn from the SDSS IV - MaNGA galaxy survey Data Release 17 (DR17). Our sample includes the whole spectrum of galaxy morphologies, and features both main-sequence and quenched objects. We discuss how different morphologies and star forming classes contribute to the number density of galaxies in the nearby Universe. We leverage our circular velocity catalogue to compute the HMF, and analyse how different galaxy morphological categories contribute to the density of matter in the nearby Universe. By combining our HMF data with previous literature computations outside our probed halo mass range, we constrain the HMF functional form between $10^{10.5}-10^{15.5}\ \rm{M_{\odot}}$.

This paper is structured as follows: In Section \ref{sec:sample and methods} we present the galaxy sample used and outline the methods employed to derive circular velocities and effective volumes. In Section \ref{sec:CVF}
we present the CVF and discuss its implications for galaxy evolution. In Section \ref{sec: Halo mass function} we show our HMF computation from IFS kinematics, constrain its functional form across 5 orders of magnitude in halo mass and compare it with the $\Lambda$CDM prediction. We provide concluding remarks in Section \ref{sec:conclusions}.

Throughout this paper, unless otherwise stated, we adopt a flat $\Lambda$CDM cosmology, with ($H_{\rm{0}}$, $\Omega_{\rm{M}}$, $\Omega_{\rm{\Lambda}}$) = (70 $\rm{km\ s^{-1}\ Mpc^{-1}}$, 0.28, 0.72). We account for the dependency of various physical parameters on Hubble's constant ($H_{\rm{0}}$) via $h_{\rm{70}}$ = $H_{\rm{0}}/(70$  $\rm{km\ s^{-1}\ Mpc^{-1}})$. Differences in assumed cosmology between such parameters and those of other data sets highlighted throughout this paper are accounted for using the prescriptions of \cite{croton_damn_2013}. In Sections \ref{sec:HMF_separate} and \ref{sec:HMF_Omega}, we switch to a Planck 2018 cosmology (\citealt{aghanim_planck_2020}) with $h_{\rm{67}}$ = $H_{\rm{0}}/(67.37$ $\rm{km\ s^{-1}\ Mpc^{-1}})$, to facilitate comparison with their estimate of the matter density in the nearby Universe.

\begin{figure*}
	\centering
	\includegraphics[width=\linewidth]{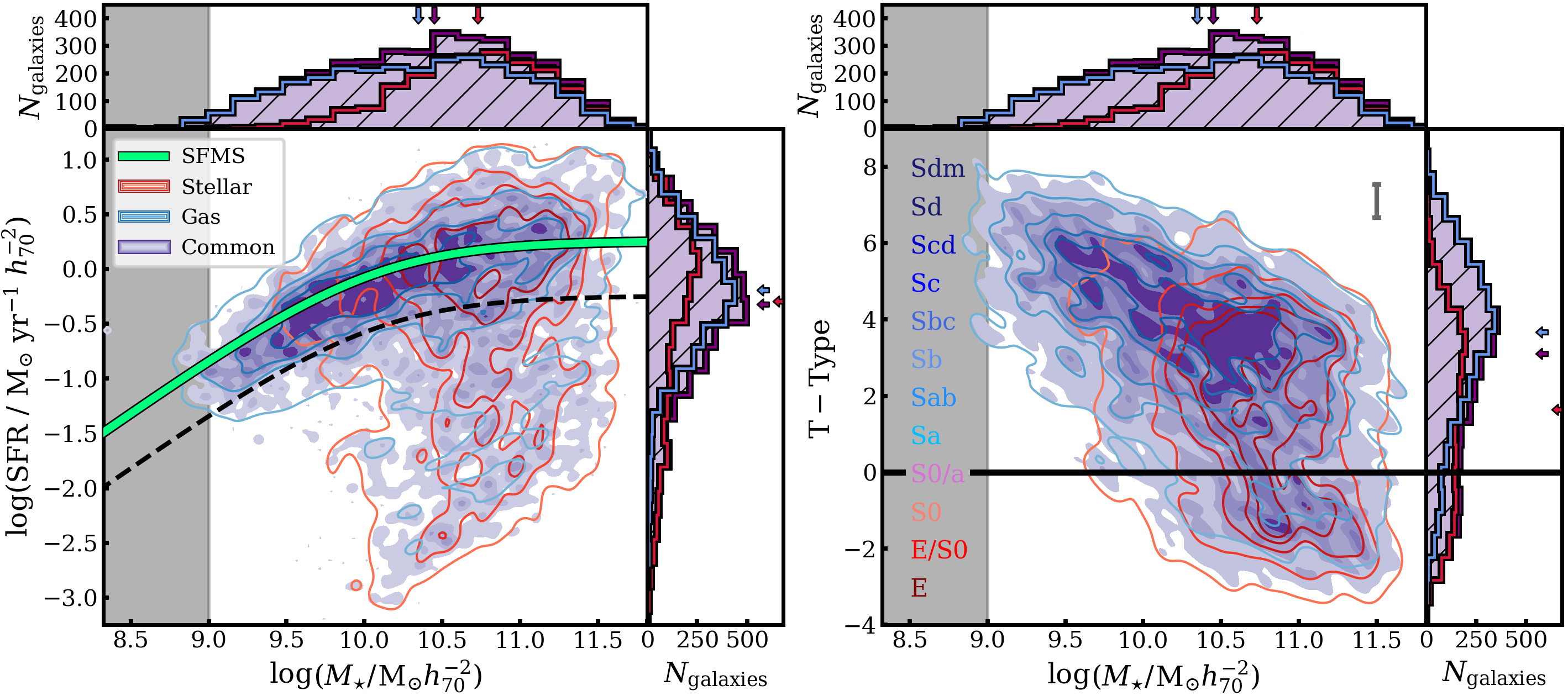}
    \caption{The galaxy samples used in this study on the SFR -$M_{\star}$ (\textbf{left}) and T-type - $M_{\star}$ (\textbf{right}) planes. The \textbf{\textcolor{red}{stellar}} and \textbf{\textcolor{blue}{gas}} samples are shown as red and blue empty isodensity contours, respectively. The \textbf{\textcolor{violet}{common}} kinematic sample (the union of the stellar and gas samples) is shown by filled purple isodensity contours. The adjacent histograms show galaxy number counts; the arrows show the median of the respective parameter's distribution, for each sample. The green curve in the left plot shows the star forming main sequence (SFMS) computed by \protect\cite{fraser-mckelvie_sami_2021} for the MaNGA sample. The morphological type corresponding to each T-type value is shown on the plot on the right. The gray error bar shows the median uncertainty in T-type for our common sample. The gray-shaded area is the region with $\log_{10}(M_{\star}/\rm{M_{\odot}})<9$, for which our common sample is statistically incomplete (see Section \ref{sec:volume weights} and Appendix \ref{sec:Appendix_completeness}).    } 
    \label{fig:sample_plot}
\end{figure*}

\section{Sample and methods}
\label{sec:sample and methods}

In this section, we present the kinematic catalogue of rotational velocities and rotational-to-dispersion velocity ($v/\sigma$) ratios (Section \ref{sec:kinematic catalogue}), the method of computing circular velocities (Section \ref{sec:circular velocities}) and galaxy effective volumes (Section \ref{sec:volume weights}).

\subsection{Kinematic catalogue and MaNGA sample}
\label{sec:kinematic catalogue}

This study uses the kinematic catalogue presented in \cite{ristea_tullyfisher_2024}, based on the final data release of the MaNGA galaxy survey (\citealt{ abdurrouf_seventeenth_2022}). 
In particular, we make use of data for the stellar and gas kinematic samples with rotation curves reaching at least 1.3 half-light radii ($R_{\rm{e}}$). Stellar masses ($M_{\star}$) and star formation rates (SFRs) used in this work are extracted from the \textit{GALEX}-Sloan-\textit{WISE} Legacy Catalogue 2 (GSWLC-2, \citealt{salim_lessigreatergalexlessigreater_2016}, \citealt{salim_dust_2018}). Galaxy T-types are available in the MaNGA Morphology Deep-Learning DR17 value added catalogue (\citealt{fischer_sdss-iv_2019}, \citealt{dominguezsanchez_sdss-iv_2021}). We exclude 103 galaxies with a misalignment between the rotation axes of stars and gas larger than 30 degrees (\citealt{ristea_sami_2022}). The gas in such objects is either being accreted or outflowing, with kinematics that are likely not tracing a state of virial equilibrium. We further exclude the 132 galaxies in our sample which do not have redshift measurements in the MaNGA Data Reduction Pipeline (DRP) summary table (\texttt{drpall\_v3\_1\_1}, compiled from the NASA Sloan Atlas catalogue, \citealt{blanton_improved_2011}). The stellar and gas samples used in this work include 2051 and 2872 galaxies, respectively. The union of these sub-samples results in a sample of 3527 galaxies (hereafter the MaNGA common kinematic sample), with a median redshift $z=0.04$. 

The samples used in this work are presented in Fig. \ref{fig:sample_plot}, on the SFR - stellar mass ($M_{\star}$) (left) and T-type - $M_{\star}$ (right) planes. The combination of the stellar and gas samples (the common kinematic sample, shown in purple) allows us to probe down to a stellar mass of $\approx 10^{8.9}\rm{M_{\odot}}$, while also sampling the quenched regime, more than 0.5 dex in $\log_{10}(\rm{SFR}/\rm{M_{\odot}}yr^{-1})$ below the star forming main sequence (SFMS) computed by \citealt{fraser-mckelvie_sami_2021}). We divide our sample based on morphology, into late-types (LTs, T-type $>$ 0) and early-types (ETs, T-type $\leqslant$ 0). We also separate our galaxies into star forming (SF, $\Delta_{\rm{SFMS}} \geqslant -0.5\ \rm{dex}$, where $\Delta_{\rm{SFMS}}$ is the offset from the SFMS) and non star forming (NSF, $\Delta_{\rm{SFMS}} < -0.5\ \rm{dex}$ ). Our common kinematic sample includes 666 ET (19 per cent of the sample) and 1139 NSF (32 per cent of the sample) galaxies spanning $\sim 1.5$ dex in $\log_{10}(M_{\star})$. The diversity of our sample in terms of morphology and SFR represents a significant improvement on previous observational studies of the CVF, largely based on \ion{H}{I} surveys (\citealt{zwaan_velocity_2010}, \citealt{papastergis_velocity_2011}), or featuring low number statistics in their ET population (\citealt{bekeraite_califa_2016}).

We make use of the stellar and gas rotational velocities at 1.3 $R_{\rm{e}}  $\ ($v_{\rm{rot}}$) and rotational-to-dispersion velocity ratios within an aperture with the same radius ($v/\sigma$), computed as described in \cite{ristea_tullyfisher_2024}. At 1.3$R_{\rm{e}}$, the rotation curve of a galaxy described by a single exponential disc component will reach its peak (\citealt{freeman_disks_1970}). In the following sub-section, we present the method of computing circular velocities using these parameters.

\begin{figure*}
	\centering
	\includegraphics[width=\linewidth]{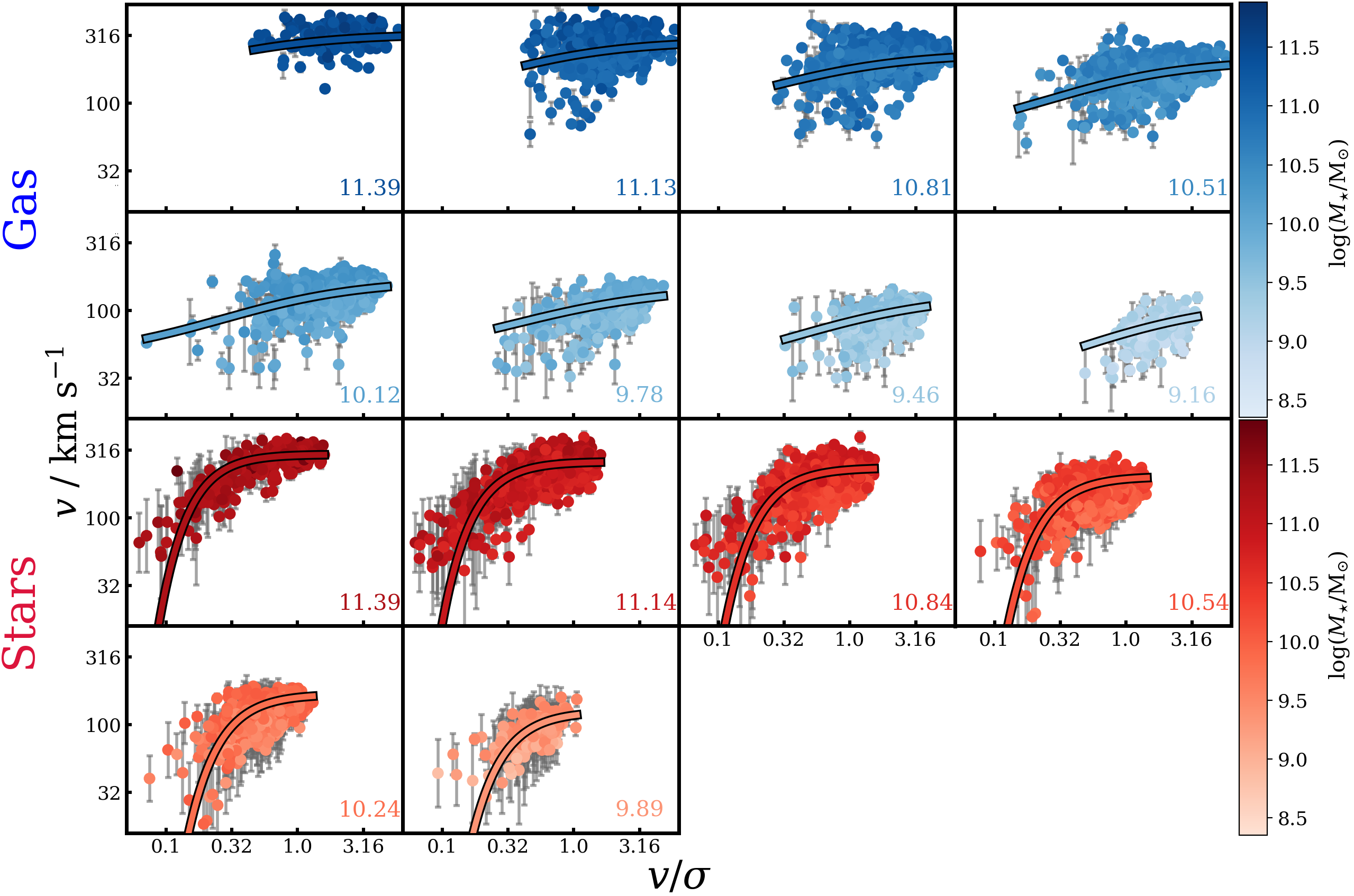}
    \caption{Kinematic description of our \textbf{\textcolor{red}{stellar}} (\textbf{left}) and \textbf{\textcolor{blue}{gas}} (\textbf{right}) samples, on the rotational velocity ($v$, measured at 1.3$R_{\rm{e}}$) - $v/\sigma$ (measured within 1.3$R_{\rm{e}}$) plane. Stellar and gas samples are split between 6 and 8 bins in $\log_{10}(M_{\star})$, respectively, as shown by the colour-coding. The curves show best-fitting sigmoid function fits (equation \ref{eq:sigmoid_2d}) to bins of $\log_{10}(M_{\star})$. The number displayed in the bottom right corner of each plot slows the median $\log_{10}(M_{\star})$ of the respective bin. The axes are linearly spaced in $\log_{10}(v/\sigma)$ and $\log_{10}(v/\rm{km\ s^{-1}})$.} 
    \label{fig:sigmoid_2d}
\end{figure*}

\subsection{Circular velocities}
\label{sec:circular velocities}

The rotational velocities calculated in \cite{ristea_tullyfisher_2024} from IFS data trace only the ordered stellar and gas rotational motion in a disc. In this work, we account for the contribution from disordered motions to the kinetic energy and compute circular velocities ($v_{\rm{circ}}$). Estimates of this parameter from IFS observations typically involve either dynamical modelling of the kinematics (\citealt{kalinova_inner_2017}, \citealt{zhu_manga_2023}) or applying a correction to the rotational velocity based on the velocity dispersion of the kinematic tracer (referred to as an asymmetric drift correction, e.g. \citealt{birkin_kaoss_2024}).

Given that approaches of computing circular velocities from stellar and gas kinematics are historically different, here we devise a consistent method applicable to both baryonic components. Furthermore, we aim to derive $v_{\rm{circ}}$ measurements using a non-parametric approach, i.e., making no assumptions of the underlying kinematic and morphologic structures of galaxies (e.g assuming the velocity ellipsoid anisotropy and/or the mass-to-light ratio, as is commonly the case for dynamical modelling approaches). 

For galaxies where the kinetic energy budget of baryons is dominated by ordered rotation (i.e., the velocity dispersion $\sigma$ is negligible compared to $v_{\rm{rot}}$), the rotational velocity is expected to be a close estimate of $v_{\rm{circ}}$, and thus a good tracer of the gravitational potential. Such galaxies have morphologies dominated by a disc component and are expected to follow the exponential disc approximation. The same assumption is underpinning studies using \ion{H}{I} line widths to calculate the circular velocity. Thus, for high-$v/\sigma$ galaxies, the velocity at 1.3$R_{\mathrm{e}}$ is expected to be consistent with that of the flat part of the rotation curve. As shown by \cite{lelli_baryonic_2019} (their Fig. 4), significant differences between the two velocity estimates (above $\sim$15 per cent) are found mostly for low-mass objects (below $\sim 10^{9}\ \mathrm{M_{\odot}}$; not present in our sample) and for massive, dispersion-dominated early-type galaxies with declining rotation curves. The latter category is representative of galaxies with low stellar $v/\sigma$ values (on the rising part of the stellar sigmoids in Fig. \ref{fig:sigmoid_2d}, bottom) and low gas fractions (and thus not present in our gas kinematic sample). 
At fixed $M_{\star}$, the vertical scatter in the stellar mass Tully-Fisher relation (i.e., the scatter in rotational velocity at fixed stellar mass) correlates strongly with the $v/\sigma$ ratio. In the nearby Universe, galaxies where the stellar and gas components are dominated by coherent rotational motions ($v/\sigma \geqslant 1$) were found to exhibit little ($<0.15\ \rm{dex}$) scatter in velocity at fixed $M_{\star}$ (\citealt{ristea_tullyfisher_2024}).

In Fig. \ref{fig:sigmoid_2d}, we plot the rotational velocities of galaxies in our stellar and gas samples at 1.3$R_{\rm{e}}$ as a function of $v/\sigma$ ratios within the same radius, split into 6 and 8 stellar mass bins, respectively. The bin widths are chosen such that each bin includes at least 20 galaxies and the bin sizes for the stellar (0.46 dex) and gas (0.44 dex) samples are approximately equal. At fixed stellar mass, galaxies appear to follow a sigmoidal trend in this parameter space regardless of the kinematic tracer. We thus elect to fit a sigmoid function to each $M_{\star}$ bin in the $\log_{10}(v)-\log_{10}(v/\sigma)$ parameter space (using least-squares regression), defined as: 

\begin{equation}
    \log_{10}(v/  \mathrm{km\ s^{-1}}  ) =      v_{\mathrm{0}} - \frac{ v_{\mathrm{0}} -  v_{\mathrm{1}}}{1+e^{-k\big(  \log_{10}(\frac{v}{\sigma}) - \log_{10}(\frac{v}{\sigma})_{\mathrm{0}} \big)}},
	\label{eq:sigmoid_2d}
\end{equation}

\noindent where $v_{0}$ and $v_1$ are the asymptotic rotational velocities at high and low $v/\sigma$, respectively (if $k<0$), or at low and high $v/\sigma$, respectively (if $k>0$), $k$ controls the sharpness of the turn between the two asymptotes and $\log_{10}(v/\sigma)_{0}$ sets the sigmoid turnover point. We list the best-fitting sigmoid parameters in each stellar mass bin in Table \ref{tab:breakdown_sigmids}.

For stellar kinematics, there is a steep increase in $v$ up to $v/\sigma \approx 0.56$ (corresponding to the separation between dynamically-cold discs and intermediate systems as determined by \citealt{fraser-mckelvie_beyond_2022}, without accounting for the effect of beam smearing; see \citealt{ristea_tullyfisher_2024}). This regime reflects a decrease in the contribution of central dispersion-dominated structures such as classical bulges to the kinetic energy of galaxies within the probed radius. At high $v/\sigma$ ($\gtrsim 1.0$), the gradient in the sigmoids approaches zero, except for the lowest stellar mass bin. This result is entirely expected due to the mass dependence of galactic rotation curves shapes (\citealt{catinella_template_2006}, \citealt{yoon_rotation_2021}) and implies that for the lowest mass galaxies in our sample, $1.3\ R_{\rm{e}}$ does not correspond to the flat part of the velocity profile.

The dynamical range of $\log_{10}(v/\sigma)$ for gas kinematics is shifted towards higher values ($5^{\rm{th}}-95^{\rm{th}}$ percentiles are 0.67 and 3.69) compared to stars, given its collisional nature. The majority ($92$ per cent) of galaxies in our sample have gas $v/\sigma \geqslant 0.56$ (not corrected for beam-smearing, which results in an observed $v/\sigma $ reduced by $\sim 24$ per cent compared to the intrinsic one; see \citealt{ristea_tullyfisher_2024}). We thus observe little rise in the gas sigmoids within the $v/\sigma$ range probed, suggesting that the ionised gas in these galaxies is found in a rotationally-supported disc, with a low contribution to the gas kinetic energy from dispersive, turbulent motions. However, the sigmoid normalisation still has a mass dependence as shown in Fig. \ref{fig:sigmoid_2d} (top).
In the case of both baryonic tracers, galaxies close to the asymptotic regime in $v$ vs. $v/\sigma$ have stellar and/or ionized gas kinematics dominated by ordered rotation.

The sigmoid fits reach their asymptotic regime at different $v/\sigma$ values for different $M_{\star}$ bins. For consistency between the two baryonic tracers, and between galaxies at different stellar masses, we choose to extrapolate all sigmoid fits out to $v/\sigma = 10$, where we measure the corresponding "asymptotic" rotational velocities. 
For the stellar kinematic sample, the sigmoids have reached their asymptotic values ($v_{0}$ in Table \ref{tab:breakdown_sigmids} and equation \ref{eq:sigmoid_2d}) by $v/\sigma = 10$. For the gas sample, the best-fitting velocities at $v/\sigma = 10$ are consistent within uncertainties with the sigmoid asymptotic values. The extrapolated velocity values from stellar and gas kinematics at fixed stellar mass (i.e., for the overlapping stellar mass bins in Table \ref{tab:breakdown_sigmids}) are consistent within 0.03 dex in all cases.

\begin{figure}
	\centering
	\includegraphics[width=\columnwidth]{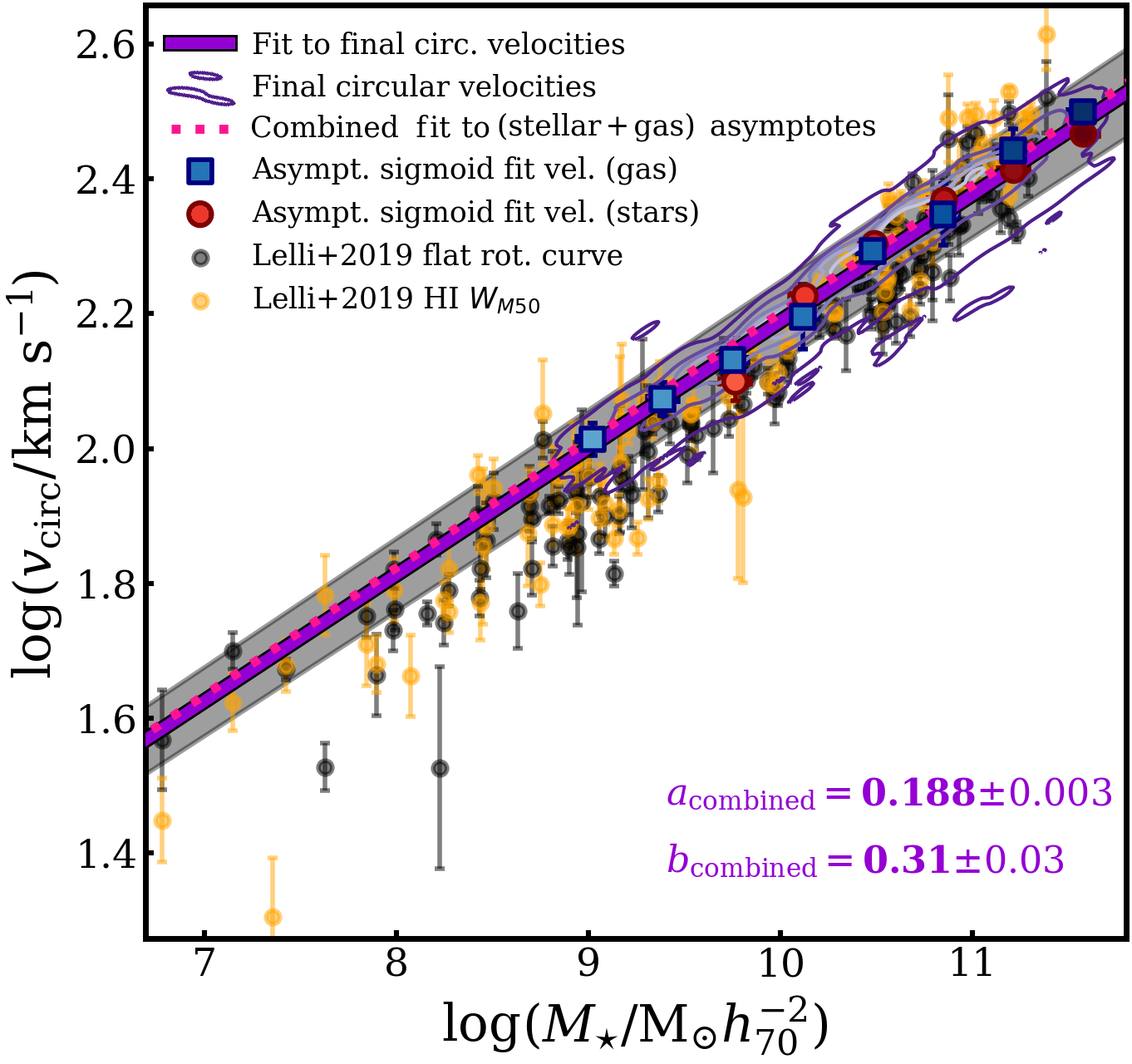}
    \caption{The circular velocity - $M_{\star}$ Tully-Fisher relation for the asymptotic sigmoid velocities outlined in Fig. \ref{fig:sigmoid_2d}, shown as blue squares and red circles for the stellar and gas samples, respectively. The colours of these symbols represent the stellar mass colour-coding from Fig. \ref{fig:sigmoid_2d}. The red circles and blue squares (asymptotic sigmoid velocities as a function of the median stellar mass in the respective bin of Fig. \ref{fig:sigmoid_2d}) were individually fitted by performing  orthogonal regression. The pink dotted line reflects the weighted mean of the two linear fits through the stellar and gas asymptotic velocities. The purple contours show the distribution of galaxies in our common kinematic sample using our final estimates of $v_{\rm{circ}}$. The solid purple line shows the best fit through these contours, with best-fitting parameters displayed. We also show the data for the SPARC galaxy sample (\citealt{lelli_sparc_2016}), coloured depending on the origin of the velocity measurement: the flat part of the H$\alpha$ rotation curve (orange) and the width of the \ion{H}{I} emission line at 50 per cent of the peak intensity (black).  } 
    \label{fig:TF_circ_vel}
\end{figure}

The dependence of the asymptotic sigmoid velocities in Fig. \ref{fig:sigmoid_2d} at $v/\sigma = 10$  on $M_{\star}$ is shown in Fig. \ref{fig:TF_circ_vel} (red circles and blue squares), compared with measurements of the Spitzer Photometry and Accurate Rotation Curves (SPARC) data set (\citealt{lelli_sparc_2016}, a set of 175 late-type and irregular galaxies with \ion{H}{I} and H$\alpha$ rotation curves). There is an agreement within uncertainties between the asymptotic measurements from stellar (red circles) and gas (blue squares) kinematics at fixed $M_{\star}$. These velocities are also largely consistent with the velocities of the SPARC galaxies, albeit being slightly higher in the lower stellar mass bin (the difference between our velocity estimate and the median velocity of the SPARC galaxies in this mass bin is 0.09 dex). This general agreement reflects the fact that circular velocities computed using our approach are tracing the virial equilibrium of galaxies. We first fit a linear model through the data for each baryonic tracer (stars and gas) separately using orthogonal linear regression implemented by the \texttt{HYPER-FIT} package (\citealt{robotham_hyper-fit_2015}). We then combine the two fits using a weighted mean (accounting for uncertainties in each fit). The combined fit is shown by the pink dotted line in Fig. \ref{fig:TF_circ_vel} ($\rm{slope}\ a=0.189\pm 0.001$ and $\rm{intercept}\ b=0.31\pm 0.05$). We first assign circular velocities to galaxies based on their $M_{\star}$ as given by this best-fitting line (hereafter referred to as $v_{\rm{circ,1}}$).

\begin{table*}
\centering
\caption{Compilation of the best-fitting parameters for the sigmoid functions (equation \ref{eq:sigmoid_2d}) in Fig. \ref{fig:sigmoid_2d}, and their uncertainties (in square brackets). The column "$\rm{N_{gal}}$" shows the number of galaxies in each bin of stellar mass. The parameters $v_{\rm{0}}$ and $v_{\rm{1}}$ are expressed in dex (reflecting the $\log_{10}$ of velocity in $\rm{km\ s^{-1}}$), while $k$ and $(v/\sigma)_{\rm{0}}$ are dimensionless. }
\setlength\tabcolsep{1.9pt}
\renewcommand{\arraystretch}{1.2}
\label{tab:breakdown_sigmids}
\begin{tabular*}{\linewidth}{@{\extracolsep{\fill}}   cccccc|cccccc}
\multicolumn{6}{c}{\textbf{\textcolor{red}{Stellar kinematics}}}                                             & \multicolumn{6}{c}{\textbf{\textcolor{blue}{Gas kinematics}}}                                                 \\
\hline
\hline
\multicolumn{1}{l}{$\log_{10}(M_{\star}/\rm{M_{\odot}})$ range} & \multicolumn{1}{l}{$\rm{N_{gal}}$} & ${v_{0}}$ & ${v_{1}}$ & $k$  & $(v/\sigma)_{\rm{0}}$ & \multicolumn{1}{l}{$\log_{10}(M_{\star}/\rm{M_{\odot}})$ range} & \multicolumn{1}{l}{$\rm{N_{gal}}$} & ${v_{0}}$ & ${v_{1}}$ & $k$ & $(v/\sigma)_{\rm{0}}$ \\
\hline
$[$11.42 - 11.88)                   & 78                       & 2.467$[4]$  &  -10.344$[5]$  & -4.99$[9]$   & -1.50$[1]$        & $[$11.44-11.88)                     & 65                       & 2.510$[7]$    & 2.14$[9]$    & -2.0$[1]$   & -0.72$[9]$         \\
$[$10.96-11.42)                     & 577    &                2.413$[3]$  &  -11.4$[1]$  &  -4.84$[6]$  & -1.49$[2]$           & $[$11.00-11.44)                     & 413                      & 2.46$[8]$    & 1.9$[1]$   & -1.95$[9]$   &  -0.68$[3]$       \\
$[$10.49-10.96)                     & 820                      &  2.368$[2]$  & -11.6$[9]$   & -4.634$[7]$   & -1.467$[2]$        & $[$10.56-11.00)                     & 696                      & 2.37$[8]$   &  1.86$[4]$  &   -1.87$[3]$ &  -0.64$[2]$        \\
$[$10.03-10.49)                     & 489                      & 2.30$[3]$   &  -12.6$[2]$  & -4.60$[5]$   & -1.45$[3]$         & $[$10.12-10.56)                     & 645                      & 2.32[9]   & 1.72[7]   & -1.868[3]  & -0.61[1]         \\
$[$9.56-10.03)                      & 141                      &  2.226$[3]$   &  -13.1$[3]$   & -4.6$[1]$   & -1.40$[2]$        & $[$9.68-10.12)                     & 581                      &  2.22[9]  & 1.65[4]   & -1.867[3]  & -0.57[2]        \\
$[$9.10-9.56)                       & 20                       & 2.01[3]   & -14.85[6]   & -4.549[6]  &  -1.398[1]       & $[$9.24-9.68)                       & 420                      & 2.17[9]   & 1.60[7]   &  -1.677[3]   & -0.50[1]        \\
                                &                          &    &    &   &         & $[$8.80-9.24)                       & 140                      &  2.13[9]  &  1.46[7]  & -1.671[3]  &  -0.47[1]       \\
                                &                          &    &    &   &         & $[$8.35-8.80)                       & 20                       &  2.1[1]  &  1.32[7]  & -1.666[3]   &  -0.44[1] \\        
\hline
\hline

\end{tabular*}
\end{table*}

\begin{figure*}
	\centering
	\includegraphics[width=\linewidth]{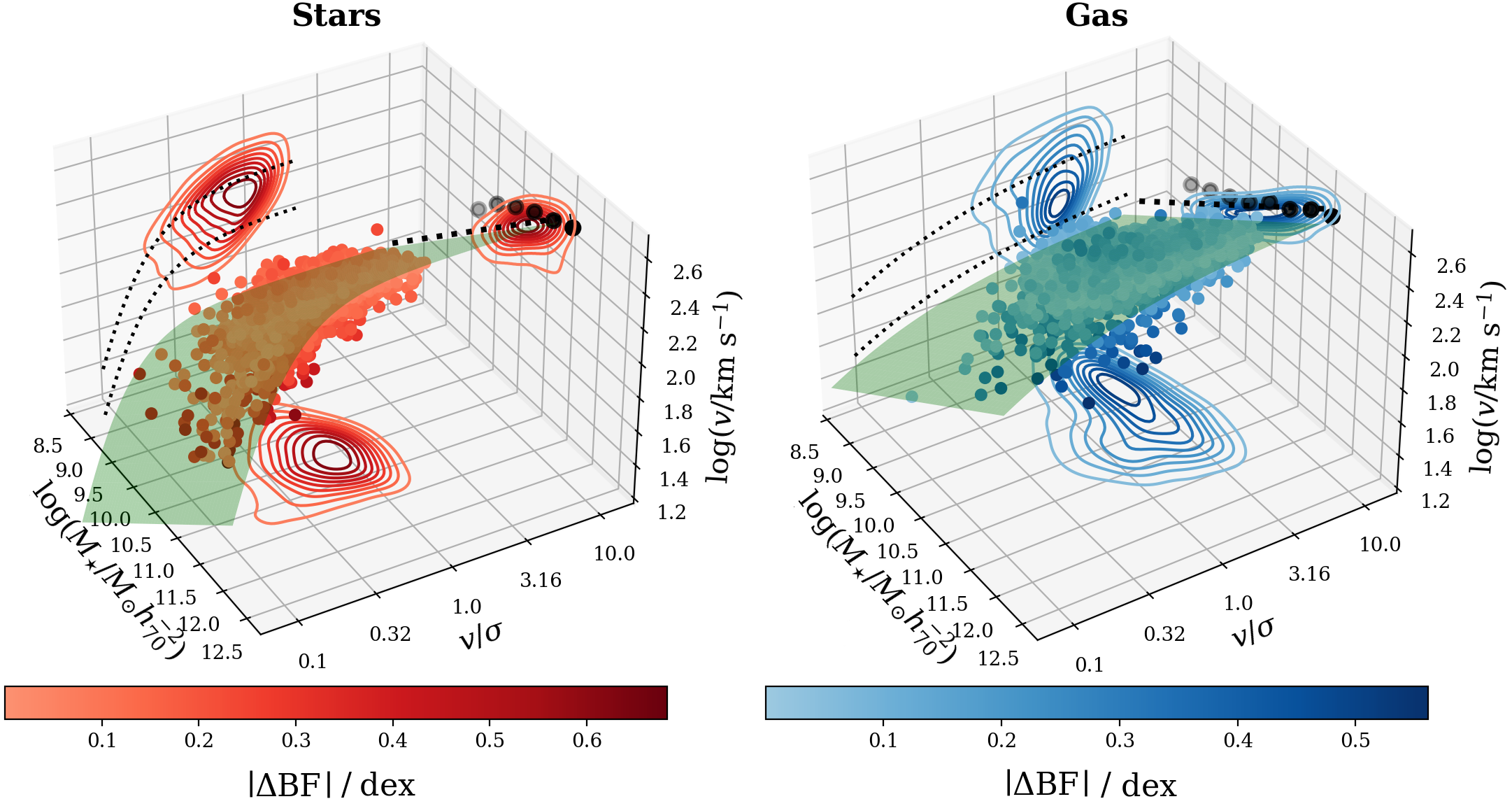}
    \caption{The \textbf{\textcolor{red}{stellar}} (\textbf{left}) and \textbf{\textcolor{blue}{gas}} (\textbf{right}) samples used in this work, shown in the $\log_{10}(M_{\star})$ - $v/\sigma$ - $\log_{10}(v)$ parameter space. The best-fitting linear-sigmoid surface to the data (equation \ref{eq:sigmoid_3d}) in each sample is shown in green. Points are colour-coded by their offset from the best-fitting surface. The distribution of the samples on each of the three planes is shown by empty isodensity contours. The dotted curves in the $\log_{10}(v)-v/\sigma$ plane show the projection of the best-fitting surface on this plane at the $16^{\rm{th}}$ and $84^{\rm{th}}$ percentiles of the stellar mass distribution. The dotted line in the  $\log_{10}(v)-M_{\star}$ plane shows the intersection of the best-fitting surface with this plane at $v/\sigma=10$. The black circles show the asymptotic values of the sigmoid fits from Fig. \ref{fig:sigmoid_2d}.     } 
    \label{fig:sigmoid_3d}
\end{figure*}

The approach presented so far has shown its utility in providing a set of reliable circular velocities corresponding to specific bins of stellar mass. However, binning galaxies in $\log_{10}(M_{\star})$ and assigning them a single circular velocity per bin means a loss of the information given by the scatter in velocity at fixed stellar mass.
There is a non-negligible scatter in $\log_{10}(v)$ at fixed $v/\sigma$ in each of the $M_{\star}$ bins of Fig. \ref{fig:sigmoid_2d} (0.04-0.12 dex). This scatter is typically higher than the median rotational velocity uncertainty in each $M_{\star}$ bin, with ratios between the two in the range 1.07-1.15. We desire to account for this scatter when computing $v_{\rm{circ}}$, given the possibility that it might be tracing the influence of galaxy evolutionary processes. To achieve this, we fit a linear-sigmoid surface to our stellar and gas samples in the $\log_{10}(M_{\star})-\log_{10}(v/\sigma)-\log_{10}(v)$ parameter space. This surface is the combination of a sigmoid function (equation \ref{eq:sigmoid_2d}) in the $\log_{10}(v/\sigma)-\log_{10}(v)$ plane and a linear function in $\log_{10}(M_{\star})-\log_{10}(v)$, defined as:

\begin{equation}
\begin{split}
    \log_{10}(v/\mathrm{km\ s^{-1}}) = v_{\mathrm{0,3D}} - \frac{ v_{\mathrm{0,3D}} -  v_{\mathrm{1,3D}}}{1+e^{-k_{\mathrm{3D}}\big(    \log_{10} \big(\frac{v}{\sigma}\big) - \log_{10} \big(\frac{v}{\sigma} \big)_{\mathrm{0,3D}} \big)}} \\ +\ A \log_{10}(M_{\star}/\mathrm{M_{\odot}}),
	\label{eq:sigmoid_3d}
\end{split} 
\end{equation}

\noindent where $A$ is the slope of the surface in the $\log_{10}(M_{\star})-\log_{10}(v)$ plane. The parameters of the sigmoidal term in equation \ref{eq:sigmoid_3d} ($v_{\rm{0,3D}}$, $v_{\rm{1,3D}}$, $k_{\rm{3D}}$, $\log_{10}(v/\sigma)_{\rm{3D}}$) have the same meaning as those in equation \ref{eq:sigmoid_2d}, with the exception that, in the case of the linear-sigmoid surface of equation \ref{eq:sigmoid_3d}, asymptotic velocities at extreme values of $v/\sigma$ are equal to the sum of either $v_{\rm{0,3D}}$ or $v_{\rm{1,3D}}$, and the $A \log_{10}(M_{\star}/\mathrm{M_{\odot}})$ term. We choose this approach for incorporating the scatter (continuous in $M_{\star}$) over a more simplistic one using the best-fitting (2-dimensional) sigmoids in Fig. \ref{fig:sigmoid_2d} given the significant width of our stellar mass bins (0.46 and 0.44 dex for the stellar and gas samples, respectively) compared to the median uncertainty in $\log_{10}(M_{\star})$ for our sample ($\approx 0.10$ dex).

The best-fitting linear-sigmoid surfaces for the stellar and gas samples are shown in Fig. \ref{fig:sigmoid_3d}, and their best-fitting parameters are displayed in Table \ref{tab:sigmoid_3d}. The intersections of the best-fitting surfaces with the $M_{\star}-\log_{10}(v)$ plane at $v/\sigma=10$ show some slight differences compared to the asymptotic velocities from Fig. \ref{fig:sigmoid_2d}. These differences are more pronounced at the lower and upper ends of the stellar masses distributions (up to 0.10 dex and 0.08 dex in $\log_{10}(v/\ \rm{km\ s^{-1}})$ for stellar and gas kinematics, respectively). The differences are due to the surface fits being driven by the region in stellar mass with the highest density of galaxies, around the median mass for each sample (see Fig. \ref{fig:sample_plot}). The median offset from the best-fitting surfaces ($|\Delta \rm{BF}_{\rm{median}}|$) is 0.11 and  0.09 dex for the stellar and gas surfaces, respectively. We add the offset of each galaxy from the best-fitting surface ($|\Delta \rm{BF}|$ in Fig. \ref{fig:sigmoid_3d}) to the asymptotic velocity values from the sigmoids in Fig. \ref{fig:sigmoid_2d} and obtain our second estimate of circular velocities, $v_{\rm{circ,2}}$ (separately for the stellar and gas samples), which incorporates the scatter in velocity. If we instead add $|\Delta \rm{BF}|$ to the asymptotic velocity values from extrapolating the linear-sigmoid surfaces in Fig. \ref{fig:sigmoid_3d} to $\log_{10}(v/\sigma)=10$, we obtain $v_{\rm{circ,2}}$ estimates consistent within uncertainties in all cases. For galaxies in our common kinematic sample with both stellar and gas velocity measurements, $v_{\rm{circ,2}}$ is calculated by averaging the measurements from the two components.

\begin{table}
\centering
\caption{ Best-fitting parameters of the linear-sigmoid surfaces in Fig. \ref{fig:sigmoid_3d} (equation \ref{eq:sigmoid_3d}) for the stellar and gas samples. }
\label{tab:sigmoid_3d}

\begin{tabular}{@{\extracolsep{\fill}}ccc}
Best-fitting parameter & \textbf{\textcolor{red}{Stellar sample}} & \textbf{\textcolor{blue}{Gas sample}} \\
\midrule
\midrule
$v_{\rm{0,3D}}$ / dex            &  $-3.8\pm 0.1$              & $-2.9\pm 0.1$            \\
$v_{\rm{1,3D}}$ / dex           &   $-1.0\pm 0.1$              & $-0.3\pm 0.1$            \\
$k_{\rm{3D}}$               &   $4.26\pm 0.02$              &   $1.38\pm 0.06$         \\
$\log_{10}(v/\sigma)_{\rm{0,3D}}$          &    $-1.29\pm 0.02$            & $-2.98\pm 0.07$            \\
$A$                  &  $0.308\pm 0.006$               &  $0.251\pm 0.004$         \\
\midrule
\midrule
\end{tabular}
\end{table}

The circular velocities used for computing the CVF (hereafter, $v_{\rm{circ}}$) are calculated by averaging the measurements from the two approaches presented ($v_{\rm{circ,1}}$ and $v_{\rm{circ,2}}$). The uncertainty in each $v_{\rm{circ}}$ measurement is taken to be half the difference between the two estimates. The "final" circular velocities are shown by the purple contours in Fig. \ref{fig:TF_circ_vel}. These $v_{\rm{circ}}$ values are available in the catalogue presented in Appendix \ref{sec:Appendix_catalogue} (Table \ref{tab:catalogue}, files available in the online version). The best-fitting circular velocity - stellar mass Tully-Fisher relation derived in this work, given by

\begin{equation}
    \log_{10}(v_{\mathrm{circ}} / \mathrm{km\ s^{-1}}) = a_{\mathrm{combined}} \times \log_{10}(M_{\star} / \mathrm{M_{\odot}} h_{70}^{-2} ) + b_{\mathrm{combined}},
\label{eq:TF}
\end{equation}

is shown by the solid purple line in Fig. \ref{fig:TF_circ_vel} (slope $a_{\mathrm{combined}} = 0.188\pm 0.003$ and intercept $b_{\mathrm{combined}} = 0.31\pm 0.03$). In the following sub-section, we present the method of computing effective volumes for the galaxies in our common kinematic sample.

\subsection{Effective volumes}
\label{sec:volume weights}

The galaxy sample observed by the MaNGA survey is not independent of luminosity; instead, it has a roughly flat $\log_{10}(M_{\star})$ distribution  (\citealt{wake_sdss-iv_2017}). Such a distribution requires a stellar mass-dependent upper redshift limit. As such, a larger volume is sampled for high-$M_{\star}$ galaxies. The MaNGA sample is thus luminosity-dependent volume-limited (Fig. 8 of \citealt{bundy_overview_2015}). 

The minimum and maximum redshifts ($z_{\rm{min}}$ and $z_{\rm{max}}$) within which a galaxy of a given absolute \textit{i}-band magnitude could have been assigned an integral-field unit (IFU) given the MaNGA selection function are publicly available through the survey's DRP summary table \texttt{drpall\_v3\_1\_1}  (\citealt{law_observing_2015}). We calculate comoving distances at the minimum and maximum redshift at which a galaxy could have been observed ($r_{\rm{min}}$ and $r_{\rm{max}}$), and determine effective volumes as: 

\begin{equation}
    V_{\mathrm{eff}}  (\mathrm{ Mpc^{3}}) = \mathcal{O}_{\mathrm{MaNGA}} (\mathrm
{sr}) \times \frac{r_{\mathrm{max}}^3 - r_{\mathrm{min}}^3}{3},
\label{eq:volumes}
\end{equation}

\noindent (\citealt{obreschkow_eddingtons_2018}). We compute the solid angle of the full MaNGA survey ($\mathcal{O}_{\rm{MaNGA}}$) by splitting the survey footprint into a grid with $1000\times1000$ bins in RA and DEC, and summing the area of grid elements which intersect MaNGA plates. This approach results in a value of $\mathcal{O}_{\rm{MaNGA}} = 2845.24\ \rm{deg^2}$.

Effective volumes calculated using equation \ref{eq:volumes} are only applicable to the \textit{entire} MaNGA sample, and not appropriate when sub-samples are used. To account for this feature, we revise the effective volumes and make them applicable to our sub-samples using the method of \cite{fraser-mckelvie_beyond_2022}. Briefly, we separate our stellar, gas and common kinematic samples into SF and NSF based on their position with respect to the SFMS (Section \ref{sec:sample and methods}). We then split our SF and NSF sub-samples in bins of stellar mass (width 0.17 dex) and redshift (width 0.0075). The number of galaxies from our kinematic samples in each bin across the $M_{\star}-z$ grid is divided by the total number of MaNGA galaxies in that respective bin. Effective volumes valid for the entire MaNGA sample are then multiplied by this completeness fraction. This process is repeated for the stellar, gas and common kinematic samples. Whenever further cuts are implemented to our common kinematic sample, this process is repeated and effective volumes are re-computed.

\begin{figure*}
	\centering
	\includegraphics[width=\linewidth]{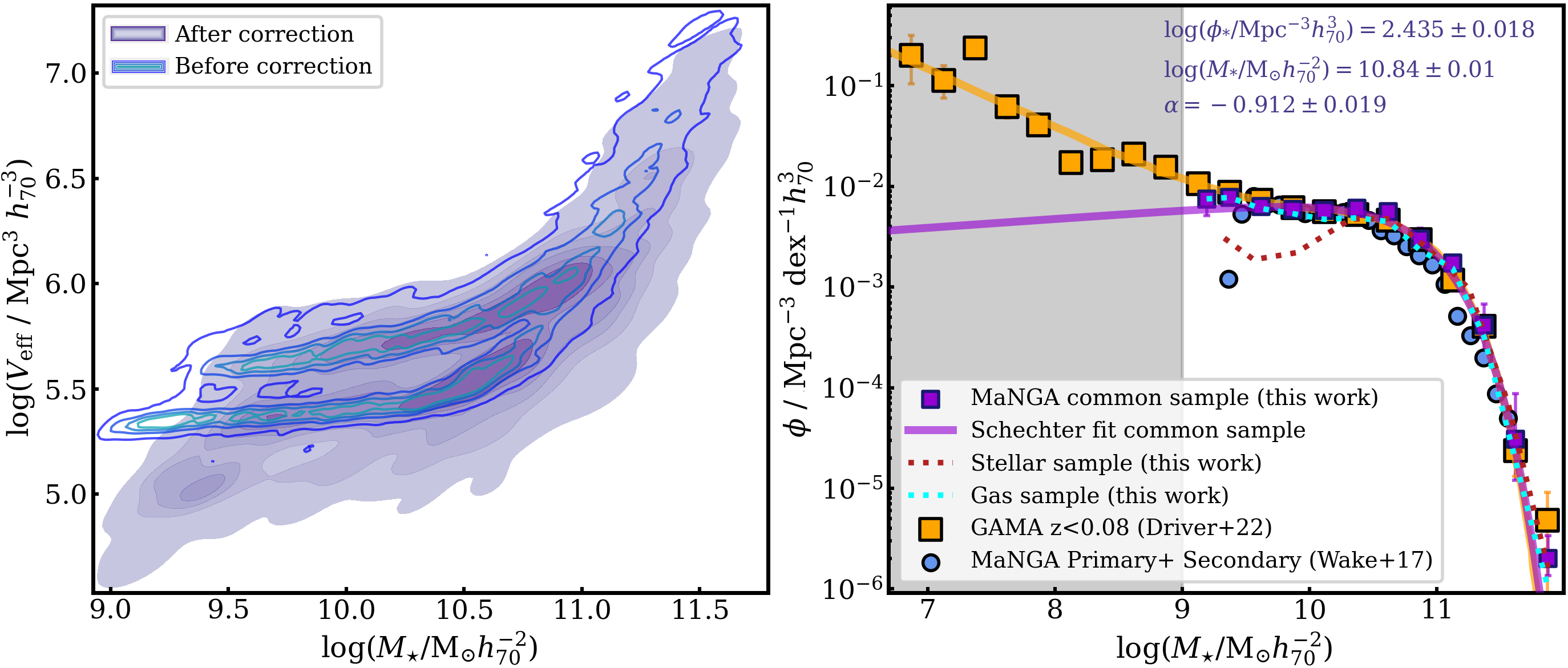}
    \caption{\textbf{Left:} The effective volumes of galaxies in our common sample as a function of stellar mass (displayed as isodensity contours), before and after the sample completeness correction with respect to the full MaNGA survey. \textbf{Right:} The stellar mass function (SMF) for our common sample shown as purple squares. Bins in $\log_{10}(M_{\star})$ are matched to those of the GAMA SMF (\citealt{driver_galaxy_2022}). The orange curve is a spline fit through the GAMA SMF. Individual SMFs for the stellar and gas samples are shown as red and light blue dotted curves, respectively. The SMF for the full MaNGA sample (\citealt{wake_sdss-iv_2017}) is shown by the blue circles. The purple solid curve shows a Schechter fit through the SMF of the common sample used in this work, with best-fitting parameters displayed in the upper right corner. The agreement between the SMF of our common sample and that of the GAMA survey is maintained down to $10^{9}\ \rm{M_{\odot}}$, below which we elect to exclude galaxies in our sample (gray-shaded area).} 
    \label{fig:MF }
\end{figure*}

Fig. \ref{fig:MF } (left) shows the result of the completeness correction for the effective volumes of our common sample. The reduction in effective volume becomes larger at lower stellar masses as we include a lower fraction of the MaNGA parent sample given our quality cuts for stellar and gas kinematics (\citealt{ristea_tullyfisher_2024}). In the right side of Fig. \ref{fig:MF } we show the stellar mass function (SMF, number density of galaxies per stellar mass bin) for our common kinematic sample (purple), in comparison with the SMF computed from the GAMA survey (\citealt{driver_galaxy_2022}) for galaxies at $z<0.08$, and the SMF for the full MaNGA sample (\citealt{wake_sdss-iv_2017}).  We find an agreement with the GAMA and full MaNGA stellar mass functions, within uncertainties, above $10^9\ \rm{M_{\odot}}$ and we elect to exclude galaxies with lower stellar masses from our sample. The SMF in Fig. \ref{fig:MF } is computed using effective volumes after implementing this cut. The agreement down to $10^9\ \rm{M_{\odot}}$ is valid for our gas sample, while the stellar one is in agreement only above $\approx 10^{10.1}\ \rm{M_{\odot}}$. For the rest of this work, we only retain galaxies in the stellar and gas samples with masses above these limits. In Appendix \ref{sec:Appendix_completeness} we present a compilation of all the samples used in this work and the respective ranges in $M_{\star}$ for which they are statistically complete. In the next section we present the CVF in the context of previous observational studies in the literature.

\section{Circular velocity function}
\label{sec:CVF}

In this section, we present the CVF derived for our MaNGA sample. We compare it to literature computations, constrain its functional form and discuss the contribution of different star forming classes and morphologies (Section \ref{sec: Circular velocity function}). We also present how different kinematic galaxy classes contribute to the number density of galaxies across the range of probed circular velocities (Sec. \ref{sec: Circular velocity function diff kinematic}).

\subsection{Circular velocity function from the MaNGA survey}
\label{sec: Circular velocity function}

\begin{figure*}
	\centering
	\includegraphics[width=\linewidth]{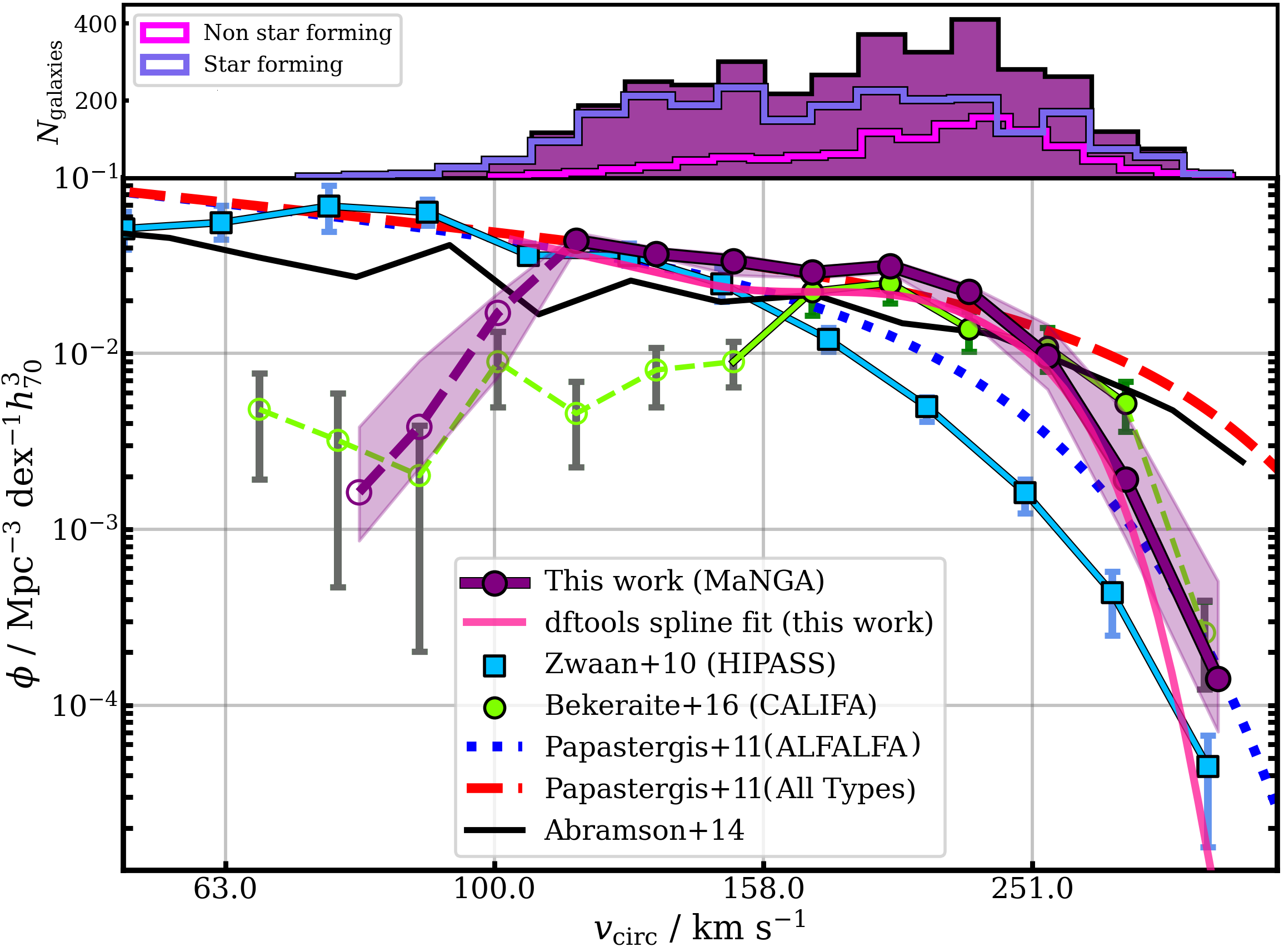}
    \caption{The circular velocity function (CVF) for our MaNGA kinematic sample (purple), in comparison with literature computations for various galaxy surveys: HIPASS (\ion{H}{I} survey; \citealt{zwaan_velocity_2010}), ALFALFA (\ion{H}{I} survey, CVF computed for 40 per cent of the full survey; \citealt{papastergis_velocity_2011}) and CALIFA (optical IFS survey, \citealt{bekeraite_califa_2016}). The computation by \protect\cite{abramson_circular_2014} is based on the SDSS survey group catalogues and employs photometric-to-kinematic scaling relations (the fundamental plane and the Tully-Fisher relation). The red dashed curve is obtained by adding the galaxy dispersion ($\sigma$) function of early-types from \protect\cite{chae_galaxy_2010} to the ALFALFA VF (blue dotted curve), assuming $v_{\rm{circ}}=\sqrt{2}\times \sigma$. The CVF of this study is binned to match that of the CALIFA survey (\citealt{bekeraite_califa_2016}). The pink curve shows a spline fit of the  MaNGA CVF, generated using the modified maximum likelihood framework implemented by \texttt{dftools}, which corrects for Eddington bias (Section \ref{sec: Circular velocity function}).  The upper panel shows the number of galaxies in our sample in bins of $v_{\rm{circ}}$, separated between star forming and non star forming objects (Section \ref{sec:kinematic catalogue}). The abscissae are linearly spaced in $\log_{10}(v_{\rm{circ}}/ \rm{km\ s^{-1}})$.  } 
    \label{fig:CVF}
\end{figure*}

\begin{figure*}
	\centering
	\includegraphics[width=\linewidth]{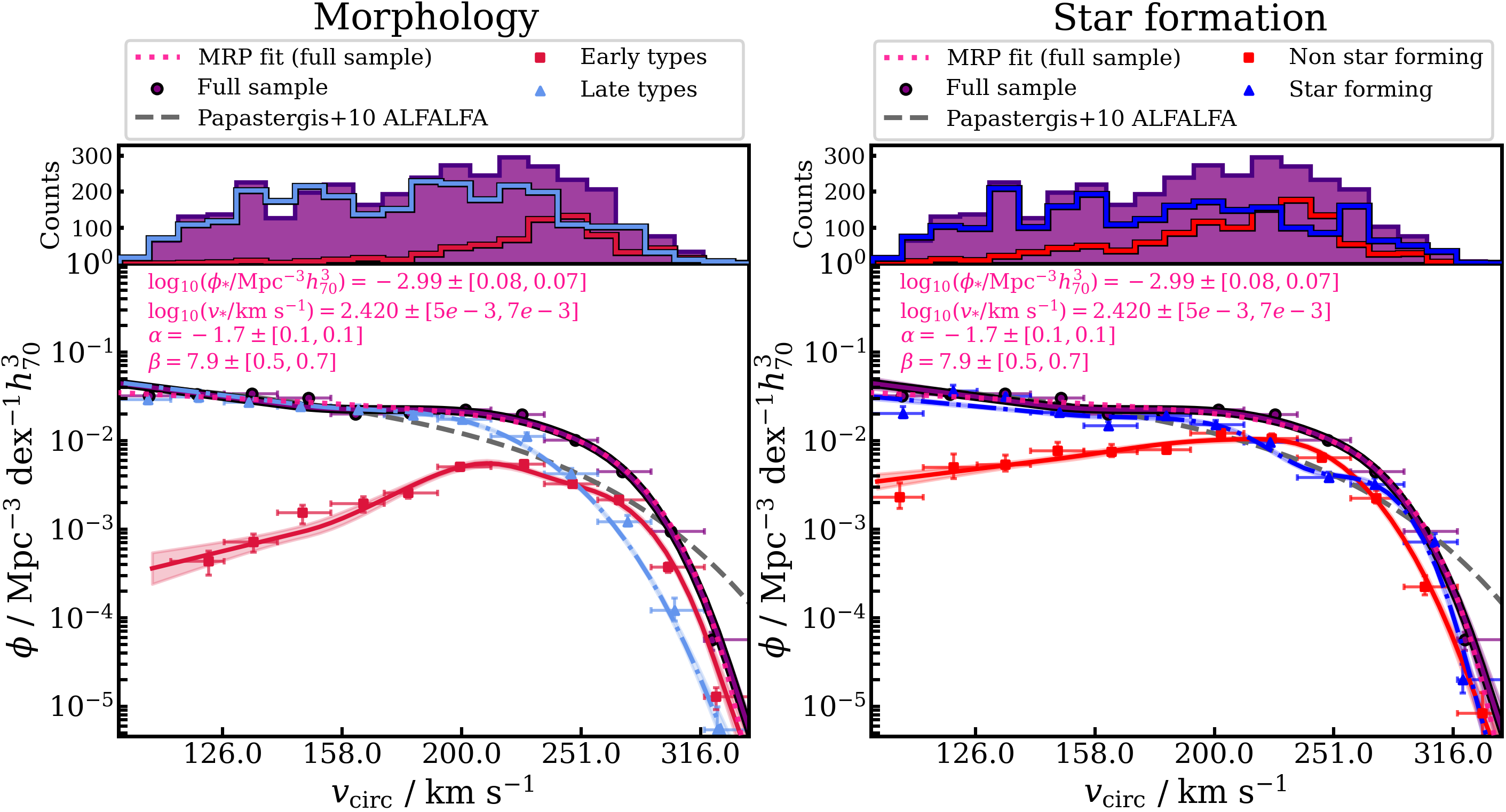}
    \caption{The circular velocity function (CVF) for our MaNGA common sample (purple) after the exclusion of galaxies with $\log_{10}(v_{\rm{circ}}/\rm{km\ s^{-1}}) \leqslant 2.011$. The purple curves show spline fits performed using \texttt{dftools}. The pink dotted curve shows the best-fitting MRP function to the CVF, with parameter values displayed. The numbers in square brackets show the upper (first) and lower (second) uncertainties in the best-fitting parameters. The upper histograms show best-fitting galaxy posterior counts based on the MML formalism implemented by \texttt{dftools}. The \textbf{left} plot shows the CVF split between the contribution to the number density from late-types and early-types. The plot on the \textbf{right} shows the contribution of star forming and non star forming galaxies to the velocity function (split based on a threshold at 0.5 dex below the SFMS). We also plot the velocity function for the ALFALFA survey at 40 per cent completeness (\citealt{papastergis_velocity_2011}), which agrees remarkably well with the CVF for star forming galaxies in our sample, below $\approx 300\ \rm{km\ s^{-1}}$.   } 
    \label{fig:CVF_morph_SF}
\end{figure*}

For a like-to-like comparison with previous studies, we first compute the CVF \textbf{$\phi$ ($\log_{10}(v_{\rm{circ}})$)} (\textbf{number of galaxies per volume}) by summing the $1/V_{\rm{eff}}$ counts corresponding to each galaxy in our common kinematic sample, in bins of $\log_{10}(v_{\rm{circ}})$. The resulting CVF is shown in Fig. \ref{fig:CVF} (purple), matched to the bins of the CALIFA CVF (\citealt{bekeraite_califa_2016}). Uncertainties in our CVF number densities reflect the number of galaxies which could belong to adjacent bins given their uncertainties in $\log_{10}(v_{\rm{circ}})$.

The MaNGA CVF shows a slightly higher number density of galaxies (a factor of $\sim 1.3-1.6$) than the CALIFA CVF within $170\ \rm{km\ s^{-1}} < v_{\rm{circ}} < 250\ \rm{km\ s^{-1}} $,  albeit marginally consistent within uncertainties. Above $250\ \rm{km\ s^{-1}}$, the MaNGA CVF exhibits a steeper drop than the CALIFA one, although still in agreement. This borderline consistency is expected due to the CALIFA CVF being computed for a sample of galaxies which includes both ETs and LTs (\citealt{bekeraite_space_2016}). The slight discrepancies can be attributed to differences in circular velocity measurement methods and kinematic morphologies probed. The numbers of ET and LT galaxies in the CALIFA sample represents $\approx 8$ and $\approx 6$ per cent of the number of galaxies with these morphologies in our common sample (Fig. \ref{fig:sample_plot}). Furthermore, the CALIFA sample predominantly includes rotationally dominated galaxies (\citealt{bekeraite_space_2016}). In contrast, our common kinematic sample includes 761 galaxies (22 per cent of our sample) classified as intermediate systems or slow rotators (using the prescription of \citealt{fraser-mckelvie_beyond_2022}; only classifiable for galaxies in the stellar kinematic sample). The computation of circular velocities for CALIFA galaxies also involves an extrapolation of rotation curves out to the optical radius $r_{\mathrm{opt}}$ (the median extent of the CALIFA sample kinematic data is $0.84r_{\mathrm{opt}}$;  \citealt{bekeraite_space_2016}), with uncertainties in velocity not reflected in Fig. \ref{fig:CVF}.

The CVF determined in this work also agrees with the ALFALFA (at 40 per cent completion, \citealt{papastergis_velocity_2011}) and HIPASS (\citealt{zwaan_velocity_2010}) computations within $100\ \rm{km\ s^{-1}} < v_{\rm{circ}} < 150\ \rm{km\ s^{-1}} $. Above $150\ \rm{km\ s^{-1}} $, the MaNGA CVF estimates a number density of galaxies $ 2.2-8.5$ times higher than the HIPASS CVF, and $ 1.5-3.0$ higher than the ALFALFA one (up to $\approx 350\ \rm{km\ s^{-1}}$, where the MaNGA and ALFALFA CVFs agree). These differences are unsurprising given that both ALFALFA and HIPASS are \ion{H}{I} surveys which omit gas-poor galaxies (\citealt{huang_arecibo_2012,obreschkow_confronting_2013}). As revealed by the upper histogram in Fig. \ref{fig:CVF}, the largest differences in number density are present in the range $\sim 170 - 250\ \rm{km\ s^{-1}} $, where the contribution of NSF galaxies to our sample is significant (42 per cent). The differences between the ALFALFA and HIPASS CVFs are due to the former having a higher sensitivity (lower detection limit), higher velocity and angular resolutions, translated to a decrease in HIPASS detections beyond 100 Mpc (\citealt{papastergis_velocity_2011}).

The CVF computed by \cite{papastergis_velocity_2011} for all galaxy types predicts a significantly higher number density of galaxies (up to a factor of 20) above $260\ \rm{km\ s^{-1}}$ than our rendition. This CVF computation considers the galaxy velocity dispersion ($\sigma$) function of ET galaxies computed by \cite{chae_galaxy_2010} and assumes a scaling of $v_{\rm{circ}} = \sqrt{2}\sigma$ (isothermal spherical halos). However, the assumption of the scaling between $v_{\rm{circ}}$ and $\sigma$ strongly affects the resulting CVF and deviations from an isothermal mass profile in ET galaxies have been widely reported from both observation-based (\citealt{dutton_kinematic_2010}) and simulation-based (\citealt{wang_early-type_2022}) studies.

A similar discrepancy is found between the MaNGA CVF and that of \cite{abramson_circular_2014} for galaxies in groups, which can be attributed to the latter making use of scaling relations (The Tully-Fisher relation and fundamental plane) to estimate circular velocities. This approach can be a potential source of systematic errors, given how these relations have been shown to exhibit large variability depending on the sample selection (e.g., the stellar mass Tully-Fisher relation, \citealt{ristea_tullyfisher_2024}). This point highlights the importance of direct kinematic measurements for a robust computation of the CVF.

The MaNGA CVF exhibits a steep drop below $\sim 100\ \rm{km \ s^{-1}}$. This effect is due to our sample not being representative of the nearby-Universe galaxy population below this threshold (i.e., an overestimation of effective volumes). Our common kinematic sample has a sharp cut in $M_{\star}$ at  $10^{9}\ \rm{M_{\odot}}$. Based on our best-fitting circular velocity Tully-Fisher relation in Fig. \ref{fig:TF_circ_vel}, this value corresponds to $\log_{10}(v_{\rm{circ}}) = 2.011$. We thus apply a cut to our common kinematic sample at this $v_{\rm{circ}}$ threshold, which excludes 71 galaxies. Effective volumes for the new sample are re-calculated as described in Section \ref{sec:volume weights}. The resulting sample is statistically complete for stellar masses $M_{\star}\geqslant10^{9.2}\ \rm{M_{\odot}}$ (see Appendix \ref{sec:Appendix_completeness}). For the remainder of this study, we only consider the common kinematic sample with $\log_{10}(v_{\rm{circ}})\geqslant 2.011$.

\begin{figure}
	\centering
	\includegraphics[width=\columnwidth]{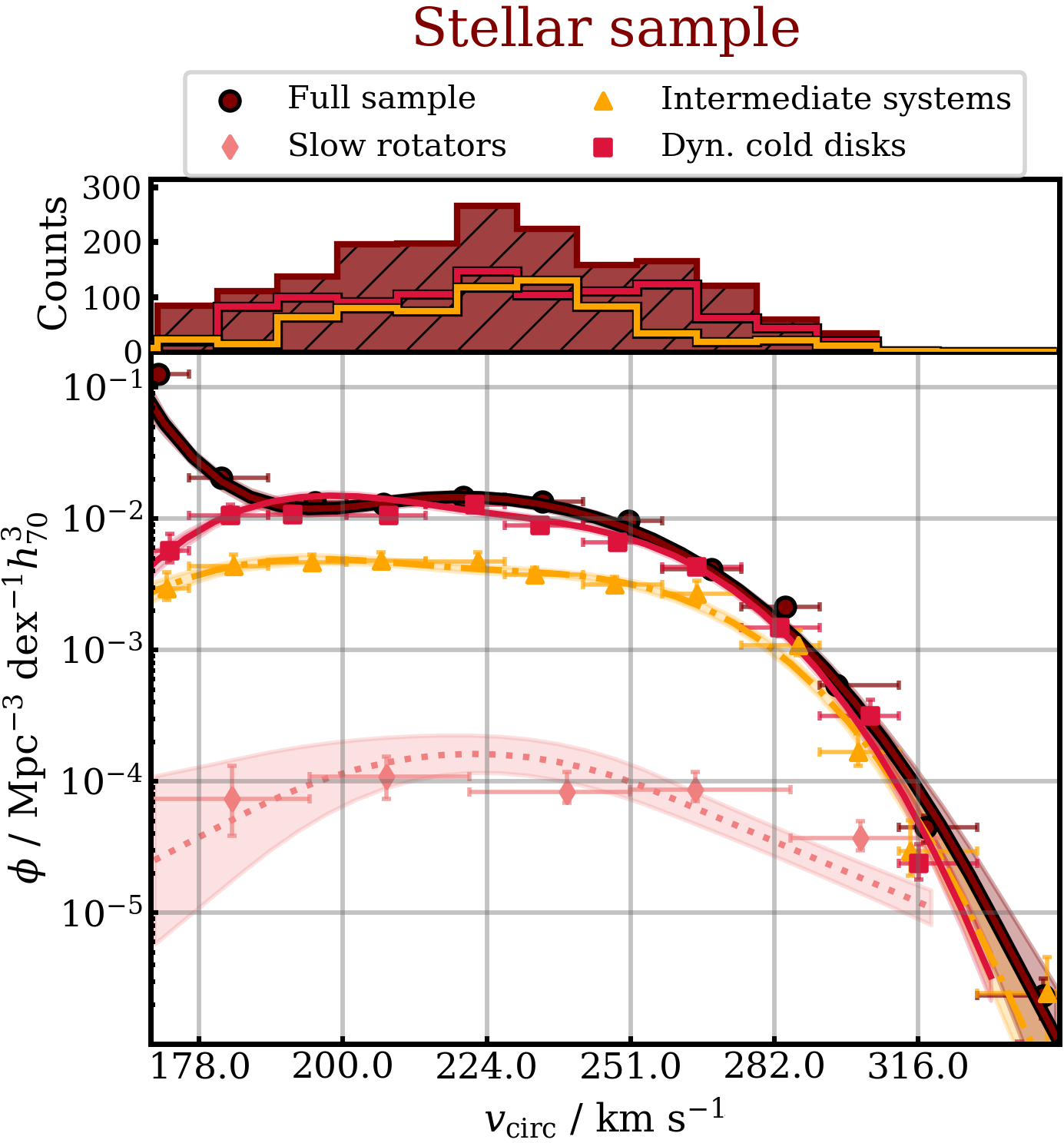}
     \caption{The circular velocity ($v_{\rm{circ}}$) functions for the full \textbf{\textcolor{red}{stellar}} (dark red) sample, generated using \texttt{dftools}. The dark red shows a spline fit through the data. The sample is split into the following kinematic classes: dynamically cold discs, intermediate systems and slow rotators (separated as described in \citealt{fraser-mckelvie_beyond_2022}). The upper histogram shows best-fitting galaxy posterior counts based on the MML formalism implemented by \texttt{dftools}. The abscissa is linearly spaced in $\log_{10}(v_{\rm{circ}})$.     } 
    \label{fig:CVF_stellar_gas}
\end{figure}

A proper determination of the empirical CVF (and indeed the generative distribution function of any galaxy parameter) from observed galaxy counts requires a formalism which: \textbf{(1)} is independent of $v_{\rm{circ}}$ binning, \textbf{(2)} robustly incorporates the observational uncertainty in $v_{\rm{circ}}$ by treating each galaxy as a free parameter with priors (probability distribution function, PDF) specified by $v_{\rm{circ}}$ and its uncertainty, and \textbf{(3)} accounts for the effect of systematic Eddington bias (\citealt{eddington_correction_1940}).
To achieve these goals, in this study we determine the CVF using the Modified Maximum Likelihood (MML) formalism implemented by the R package \texttt{dftools} (\citealt{obreschkow_eddingtons_2018}). This Bayesian MML method involves an algorithm that iteratively solves a maximum likelihood estimator and then updates the data (galaxy counts in $v_{\rm{circ}}$) based on the previous fit and on the errors. Input uncertainties in $v_{\rm{circ}}$ are determined as specified in Section \ref{sec:circular velocities}. 

The correction for Eddington bias can be summarised as follows: a steep CVF varies considerably across the PDF of an uncertain $v_{\rm{circ}}$ measurement (a normal distribution centred on $v_{\rm{circ}}$ with standard deviation equal to the error in $v_{\rm{circ}}$). In such a case, the mode of this PDF is not a good estimate of the true $\log_{10}(v_{\rm{circ}})$. A bias-corrected PDF is used instead, which takes into account the steepness of the best-fitting CVF (see equation 6 of \citealt{obreschkow_eddingtons_2018}). The magnitude of the Eddington bias is highlighted in Fig. \ref{fig:CVF}, where we also plot a spline fit of the MaNGA CVF after applying this formalism. Accounting for Eddington bias results in a shift of the CVF to lower $v_{\rm{circ}}$, more pronounced in the steep, high-velocity end.

In the following, we use the \texttt{dffit} routine of \texttt{dftools} to construct a binned CVF(bias-corrected) and model it using spline interpolation. We also elect to fit an MRP function (a modified Schechter function where the sharpness of the exponential cut-off can be controlled; \citealt{murray_empirical_2018}) given by:

\begin{equation}
\begin{split}
    \phi\  /\ \mathrm{Mpc^{-3}}\mathrm{dex^{-1}} = \frac{\mathrm{d} \textit{N}}{\mathrm{d}V\ d[\log_{10}(    v_{\mathrm{circ}}/\mathrm{km\ s^{-1}}  ) ] } \\ =  \ln(10) \phi_{*}\beta \bigg(\frac{v_{\mathrm{circ}}}{v_{*}} \bigg)^{\alpha +1} \mathrm{exp} \bigg[ {-\big(\frac{v_{\mathrm{circ}}}{v_{*}} \big)^{\beta}  } \bigg],
	\label{eq:MRP}
\end{split}
\end{equation}

\noindent where $\phi_{*}$ is the number density normalisation, $v_{*}$ is the normalisation in circular velocity (the circular velocity at the turn-over point), $\alpha$ is the faint-end slope and $\beta$ is the exponential softening parameter at high $v_{\rm{circ}}$. This function has been shown to be an accurate representation of the HMF, which is expected to be closely related to the CVF under the assumption that $v_{\rm{circ }}$ is tracing the gravitational potential of dark-matter halos. We use $\rm{N}=10^{4}$ bootstrapping iterations to estimate the covariances of the best-fitting parameters. 

The spline and MRP fits to our CVF are shown in Fig. \ref{fig:CVF_morph_SF}. The upper histograms show best-fitting posterior galaxy counts in bins of $v_{\rm{circ}}$, after accounting for Eddington bias. The CVF for our entire common sample is very well described by an MRP function, with best fitting parameters: $\log_{10}(\phi_{*}/\mathrm{Mpc^{-3}}h_{70}^3)=-2.99 \substack{+0.08 \\ -0.07}$;  $\log_{10}(v_{*}/\rm{km\ s^{-1}})=2.420 \substack{+0.005 \\ -0.007}$; $\alpha = -1.7 \pm 0.1$; $\beta = 7.9 \substack{+0.5 \\ -0.7}$. The MRP-like behavior demonstrates the fact that the circular velocities computed in this work are a tracer of the total halo mass of a galaxy.   

The left panel of Fig. \ref{fig:CVF_morph_SF} shows the contribution to the number density from ET and LT galaxies. We find that LTs dominate the number density of galaxies in the nearby Universe up to $v_{\rm{circ}} = 250\ \rm{km\ s^{-1}}$ while in the high circular velocity regime ($250-350\ \rm{km\ s^{-1}}$), ETs are dominant. In the right panel of Fig. \ref{fig:CVF_morph_SF}, we show the CVF split by contributions from SF and NSF galaxies. The SF population dominates the number density of galaxies in the nearby Universe below $200\ \rm{km\ s^{-1}}$. Above this value, the SF and NSF classes have number densities consistent within uncertainties. The CVF of SF galaxies agrees remarkably well with the ALFALFA one (\citealt{papastergis_velocity_2011}) within $100\ \rm{km\ s^{-1}} \leqslant v_{\rm{circ}} \leqslant 300 \ \rm{km\ s^{-1}}$ (the ratio of ALFALFA to MaNGA SF CVF number densities is between 0.84-1.21 in this range). At higher velocities, the MaNGA SF CVF decreases steeper than the ALFALFA one, a discrepancy that can be attributed largely to Eddington bias, for which the MaNGA CVF is corrected. 

In summary, we constructed a CVF for a sample of galaxies of all morphological types, representative of the nearby-Universe galaxy population within $10^{9.2}\leqslant M_{\star}/\rm{M_{\odot}}\leqslant 10^{11.9}$. We find slightly higher number densities that those of the CALIFA CVF (albeit still marginally consistent within uncertainties), and significantly higher number densities in the high-velocity end than previously reported  in studies based on \ion{H}{I} surveys (Fig. \ref{fig:CVF}). We robustly account for measurement errors in $v_{\rm{circ}}$ and for Eddington bias, finding that the CVF in the nearby Universe is well described by an MRP function (Fig. \ref{fig:CVF_morph_SF}). LTs dominate the number density of galaxies for $v_{\rm{circ}}\leqslant250\ \rm{km\ s^{-1}}$, above which ETs are dominant. Similarly, SF objects dominate the density of galaxies below $200\ \rm{km\ s^{-1}}$, above which SF and NSF populations contribute similarly to the nearby Universe galaxy population.

\subsection{Circular velocity function - contribution from different kinematic classes}
\label{sec: Circular velocity function diff kinematic}

We leverage the availability of $v/\sigma$ measurements for stellar kinematics and study how different kinematic classes contribute to the nearby-Universe CVF. We perform this analysis only for the stellar kinematic sample, as shown in Fig. \ref{fig:CVF_stellar_gas}, with the mention that this sample is only representative of the nearby-Universe galaxy population above $10^{10.1}\ \rm{M_{\odot}}$ (see Table \ref{tab:samples_completeness}).

We split the stellar sample into slow and fast rotators based on the delimitation of \cite{cappellari_structure_2016} (their equation 19). Fast rotators are subsequently divided into dynamically cold discs ($v/\sigma \geqslant 0.56$, not corrected for beam-smearing; see \citealt{ristea_tullyfisher_2024}) and intermediate systems ($v/\sigma < 0.56$), as proposed by \cite{fraser-mckelvie_beyond_2022}. We find that dynamically cold discs are dominating the number density of galaxies in the range $170\ \rm{km\ s^{-1}} \leqslant v_{\rm{circ}} \leqslant 315\ \rm{km\ s^{-1}}$. In the high-velocity end, intermediate systems and dynamically cold discs have comparable number densities, within uncertainties. The CVF for slow rotators shows little variations across the range of probed velocities, with only a slight decrease in number density at velocities above $\sim 280 \rm{km\ s^{-1}}$ (by a factor of $\approx 2$).

Slow rotators contribute little to the number density of galaxies across the whole range of probed $v_{\rm{circ}}$ (from 1 per cent at $v_{\rm{circ}}=180\ \rm{km\ s^{-1}}$ to 11 per cent at $v_{\rm{circ}}=300\ \rm{km\ s^{-1}}$). These results are in qualitative agreement with the findings of \cite{rigamonti_bang-manga_2024}, \cite{fraser-mckelvie_beyond_2022} and \cite{guo_sami_2020}, who found similar trends with stellar mass which resulted in a contribution from slow rotators of 8-10 per cent to the mass budget of galaxies in the nearby Universe, in the mass range $10^{9.75}<M_{\star}/M_{\odot}<10^{11.75}$. Our results extend these findings to circular velocity, a proxy for halo mass. These findings show that the fractional contribution of slow rotators to the number density of galaxies in the nearby Universe increases with circular velocity (halo mass), due to a decrease in the number density of intermediate systems and cold discs, while the number of slow rotators per unit volume remains approximately constant.

In summary, our results show a dominance from dynamically cold discs to the number density of nearby-Universe galaxies in the in the range $170-315\ \rm{km\ s^{-1}}$. In the high velocity end, the contribution of intermediate systems becomes significant, while slow rotators contribute little to the number density (up to 11 per cent) across the whole range of probed circular velocities. Our results are in qualitative agreement with previous literature examining these trends with stellar mass, and are expanding these findings to the circular velocity, a proxy for the potential well of galaxies.

\begin{figure}
	\centering
	\includegraphics[width=\columnwidth]{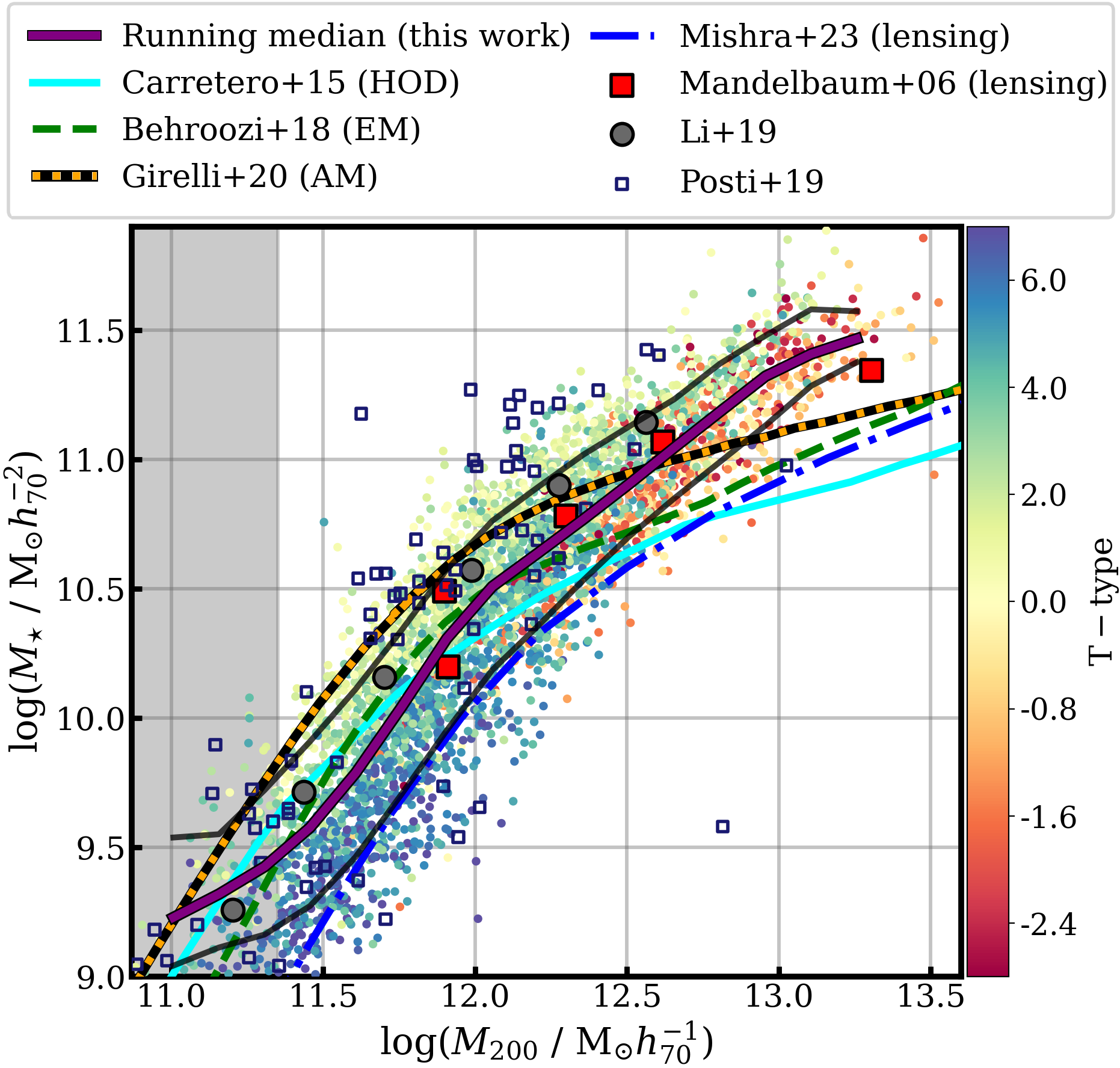}
    \caption{ The stellar mass ($M_{\star}$) - halo mass ($M_{200}$) relation for our MaNGA common sample (after the exclusion of galaxies with $\log_{10}(v_{\rm{circ}}/\rm{km\ s^{-1}}) \leqslant 2.011$), colour-coded by their T-type. The purple solid curve shows the running median through our data, and the black solid curves reflect the $16^{\rm{th}}$ and $84^{\rm{th}}$ percentiles. Below $\log_{10}(M_{200}/\rm{M_{\odot}}) \cong 11.35$ (gray-shaded area), the running median has a reduced gradient compared to higher $M_{200}$ due to our previous cut at $\log_{10}(M_{\star}/\rm{M_{\odot}}) = 9$.  We plot several literature computations of this relation, based on: halo occupation density (HOD, \citealt{carretero_algorithm_2015}), empirical modeling (EM, \citealt{behroozi_universemachine_2019}), abundance matching (AM, \citealt{girelli_stellar--halo_2020}), galaxy lensing (\citealt{mandelbaum_galaxy_2006}, \citealt{mishra_stellar_2023}) and rotation curve modelling with a dark-matter halo component (\citealt{li_halo_2019}, assuming an abundance matching SMHM prior and an NFW profile; \citealt{posti_peak_2019}).} 
    \label{fig:stellar_mass_halo_mass}
\end{figure}

\begin{figure*}
	\centering
	\includegraphics[width=\linewidth]{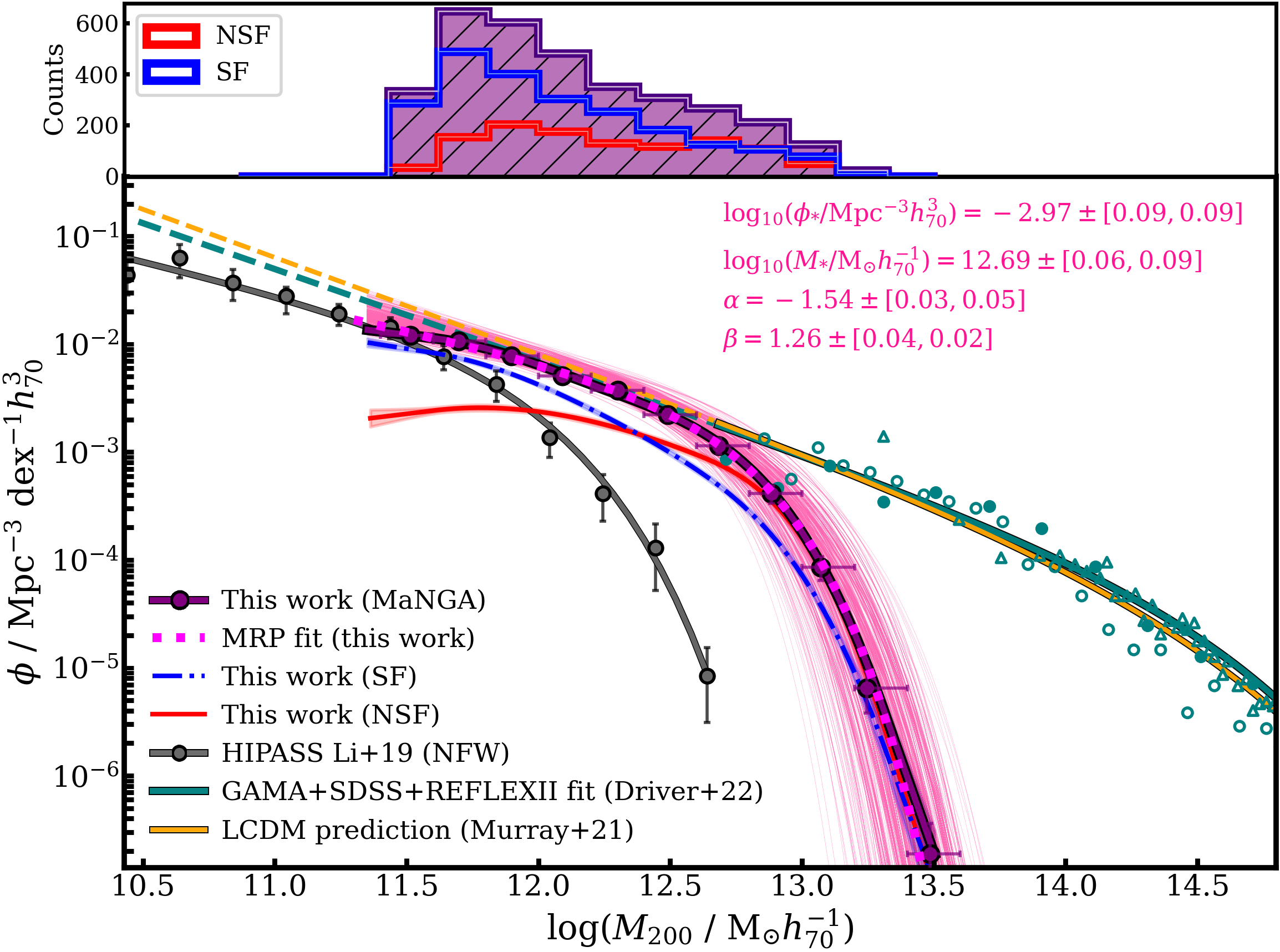}
    \caption{The halo mass function (HMF) for our MaNGA common sample (purple circles) after the exclusion of galaxies with $\log_{10}(M_{\rm{200}}/\rm{M_{\odot}})\leqslant 11.35$. The solid purple curve shows a spline interpolation, and the pink dotted curve is the best-fitting MRP function, with parameters displayed (computed using \texttt{dftools}). The numbers in square brackets show the upper (first) and lower (second) uncertainties in the best-fitting parameters. The thin pink curves show the result of scattering MaNGA HMF data points about their uncertainties in $M_{200}$ (assuming a lognormal distribution) and re-performing the fit (repeated 1000 times). The contributions to the HMF from star forming and non star forming galaxy sub-samples are shown in blue and red, respectively. The upper histogram shows best-fitting galaxy posterior counts. We also plot the best-fitting HMF to the GAMA (filled green circles), SDSS (empty green circles) and RELFEXII (empty green triangles) data from \protect \cite{driver_empirical_2022}, reflecting number densities of galaxy groups and clusters (the solid green curve is showing the best-fitting MRP function to this data, displayed as dashed below the lower mass limit of the data). The $\Lambda$CDM prediction of the HMF from \protect\cite{murray_thehalomod_2021} is shown in orange. The HMF for the HIPASS survey computed by \protect\cite{li_halo_2019} (assuming an NFW profile) is shown by gray circles, with the best-fitting double-Schechter function shown by the gray curve. } 
    \label{fig:HMF}
\end{figure*}

\section{Halo mass function}
\label{sec: Halo mass function}

In this section, we extend our analysis of the CVF to the Halo mass function (HMF). We combine our findings with HMFs from the literature and constrain this metric across $\sim$5 orders of magnitude in halo mass. In the next sub-section, we present the process of determining dark matter halo sizes, $R_{200}$ (the radius within which the average density is 200 times the Universe's critical density) for our galaxy sample.

\subsection{Stellar mass - halo mass relation}
\label{sec: SMHM rel}

We aim to calculate dark matter halo radii ($R_{200}$) for our common galaxy sample using the galaxy-halo size ($R_{\rm{e}}-R_{200}$) relation. This relation has been found to exhibit significant scatter which correlates with galactic morphology (\citealt{huang_relations_2017}) such that ETs have larger halos compared to LTs, at fixed $R_{\rm{e}}$. 

In this work, we use the $R_{\rm{e}} - R_{200}$ relation of \cite{lapi_precision_2018}, calibrated for disc morphologies. We only apply this scaling to galaxies in our common sample with similar morphologies (LTs, T-type $ > 0$; see Fig. \ref{fig:sample_plot}). For galaxies with T-type $ \leqslant 0$ (ETs), we use the relation derived by \cite{kravtsov_size-virial_2013} for a sample of various morphologies, but dominated by ETs. 

We calculate the halo mass within $R_{200}$, $M_{200}$, for our common kinematic sample as:

\begin{equation}
    M_{200} = \frac{R_{200}v_{\mathrm{circ}}^2  }{G},
    \label{eq:m200}
\end{equation}

\noindent where $G = 6.67 \ \times 10^{-11}\ \rm{N\ m^{2}\ kg^{-2}}$ is the gravitational constant (see e.g \citealt{navarro_structure_1996}, \citealt{navarro_inner_2004}). The halo mass $M_{200}$ can however be estimated using only $R_{200}$, the Hubble constant $H(z)$ and the gravitational constant $G$, or conversely using only the circular velocity $v_{\rm{circ}} \equiv v_{200}$ and the same two constants (see \citealt{mo_formation_1998}). While in this work we choose the method in equation \ref{eq:m200} due to its reliability in reproducing literature SMHM relations (Fig. \ref{fig:stellar_mass_halo_mass}), we highlight the significant spread in $M_{200}$ values obtained from different analytical relations (using either $R_{200}$ or $v_{\rm{circ}}$), which represents a source of systematic uncertainty. We discuss the magnitude of this systematic uncertainty, and its impact on the HMF in Appendix \ref{sec:Appendix_errors}.

Halo masses and their uncertainties computed in this work are available in the catalogue presented in Appendix \ref{sec:Appendix_catalogue} (Table \ref{tab:catalogue}, files available in the online version). Uncertainties in $M_{200}$ are computed by propagating the errors in $v_{\rm{circ}}$ and assuming a constant uncertainty in $R_{200}$ equal to the median scatter in the $\log_{10}(R_{\rm{e}})-\log_{10}(R_{200})$ relations employed, at fixed galaxy size (0.05 dex). Halo mass uncertainties are dominated by errors in $v_{\rm{circ}}$ and robust to the choice of the $R_{200}$ uncertainty. To test the suitability of this method, we plot the stellar mass - halo mass (SMHM) relation in Fig. \ref{fig:stellar_mass_halo_mass}, colour-coded by T-type. We overplot several literature computations of the SMHM relation, determined using several approaches: lensing data (\citealt{mandelbaum_galaxy_2006}, \citealt{mishra_stellar_2023}), halo occupation densities (HOD, \citealt{carretero_algorithm_2015}), empirical modelling (EM, \citealt{behroozi_universemachine_2019}) and abundance matching (AM, \citealt{girelli_stellar--halo_2020}). The data from \citealt{li_halo_2019} reflect galaxies in the HIPASS survey, for which $M_{200}$ was computed based on a scaling relation between this parameter and the \ion{H}{I} line width at 50 per cent of the peak flux density, determined by modelling the rotation curves of galaxies in the SPARC data set with a dark matter halo component (assuming a Navaro-Frenk-White, NFW profile; \citealt{navarro_structure_1996}), and imposing a prior based on a SMHM relation from abundance matching. The $M_{\star}-M_{200}$ relation for individual galaxies in a sub-set of the SPARC data set are also shown (\citealt{posti_peak_2019}, halo masses determined from dynamical modelling of \ion{H}{I} rotation curves, in conjunction with \ion{H}{I} and 3.6$\mu \mathrm{m}$ intensity maps).

The running median for our common kinematic sample in the $M_{\star} - M_{200}$ parameter space agrees with literature computation, within the $16^{\rm{th}}-84^{\rm{th}}$ percentiles, up to $M_{200}\approx10^{12.7}\rm{M_{\odot}}$. There are, however, significant differences between SMHM relations computed using different approaches. Although a comprehensive comparison between different SMHM determination methods is beyond our scope, we refer the reader to \cite{girelli_stellar--halo_2020} for such a discussion.

The combination of the two $R_{\rm{e}}-R_{200}$ scaling relations used in this work for LTs (\citealt{lapi_precision_2018}) and ETs (\citealt{kravtsov_size-virial_2013}) manages to bring all morphological types to a SMHM relation with scatter $\approx 0.18$ dex at fixed $M_{\star}$. This relation for our common sample exhibits a steeper gradient at $M_{200}\geqslant 10^{12.7}\rm{M_{\odot}}$ than those of \cite{carretero_algorithm_2015}, \citealt{behroozi_universemachine_2019}, \citealt{girelli_stellar--halo_2020} and \citealt{mishra_stellar_2023}. This difference is because our $M_{200}$ values are representative of the masses enclosed by halos of individual galaxies. For galaxies which are part of groups in our sample, their halo masses represent the contribution of the respective galaxy to the total mass of the group. In contrast, halo masses in the high-$M_{200}$ regime specific to the literature computations mentioned represent the total masses of group and cluster halos (rather than the contribution of individual galaxies in the group/cluster to these masses), with $M_{\star}$ representing the stellar mass of the group's central galaxy. Our approach results in a scatter in the SMHM relation that does not exceed the spread of literature computations at fixed stellar mass. This comparison gives confidence that the method used here provides a reliable qualitative description of the HMF in the mass regime probed.

Below $M_{200}=10^{11.35} \rm{M_{\odot}}$, the SMHM relation for our common sample exhibits a flatter gradient compared to literature computations. This feature is due to our lower $M_{\star}$ cut at $10^{9}\rm{M_{\odot}}$, and as a result our sample exhibits a lack of statistical completeness in this range. We thus elect to exclude galaxies with $M_{200}<10^{11.35}\ \rm{M_{\odot}}$ from our common kinematic sample for the remainder of this analysis. Effective volumes for this new sample are re-calculated as described in Section \ref{sec:volume weights}. This cut excludes 78 galaxies and results in a sample statistically complete between $ 10^{9.2}\rm{M_{\odot}} \leqslant M_{\star} \leqslant 10^{11.9}\rm{M_{\odot}}$ (Table \ref{tab:samples_completeness}). 
In the next section we present the HMF for our  galaxy sample.

\subsection{Halo mass function - constraints from IFS kinematics}
\label{sec:HMF_IFS}

The HMF for our common kinematic sample with $M_{200}\geqslant\ 10^{11.35} \rm{M_{\odot}}$ is shown in Fig. \ref{fig:HMF}. The dark purple curve shows a spline interpolation while the pink dotted curve is the best-fitting MRP function (equation \ref{eq:MRP}), computed through the MML formalism implemented by \texttt{dffit}. Best-fitting parameter covariances are estimated using $\rm{N}=10^{4}$ bootstrapping iterations. As detailed in Appendix \ref{sec:Appendix_errors} (Table \ref{tab:systematic_effect}), the best-fitting parameters of the MaNGA HMF change by up to 8 per cent when systematic uncertainties in $M_{200}$ are considered. We also plot the HMF of \cite{li_halo_2019} for the HIPASS Survey, with the mention that the $M_{200}$ values in this study reflect only the mass of dark matter and do not include baryons. Based on the HIPASS SMHM relation computed by \cite{li_halo_2019} (Fig. \ref{fig:stellar_mass_halo_mass}), $M_{\star}$ is between 1-4 per cent of $M_{200}$ for this sample. Using the neutral and molecular gas main sequences of \cite{feldmann_link_2020} for nearby galaxies, the ratio of baryonic mass (stellar, \ion{H}{I} and $\rm{H_2}$ gas masses) to $M_{200}$ is in the range 2-5 per cent for the HIPASS sample in \cite{li_halo_2019}. The HMF of \cite{driver_empirical_2022} for galaxy groups and clusters, combining data from the GAMA, SDSS (data release 12) and REFLEX II surveys is also displayed, adjusted to a median redshift $\tilde{z} = 0$. Finally, we show the $\Lambda$CDM-based prediction of \cite{murray_thehalomod_2021} (also adjusted to $\tilde{z} = 0$). The data of the CVF and HMF determined in this work are shown in Table \ref{tab:vf_hmf}, together with their uncertainties.

\begin{table*}
\centering
\caption{Circular velocity function (\textbf{left column}) and halo mass function (\textbf{right column}) data derived in this paper, and their associated uncertainties (Fig. \ref{fig:CVF_morph_SF} and Fig. \ref{fig:HMF}, respectively - purple data points). The HMF data derived in this work and used in the joint fit of Fig. \ref{fig:HMF_Integral} are displayed in the upper six rows. To convert the HMF data to the values displayed in Fig. \ref{fig:HMF_Integral}, one must convert to the Planck 2018 cosmology, i.e. multiply $M_{200}$ by 1.039, and divide $\phi$ by  1.121.    }
\label{tab:vf_hmf}
\setlength{\tabcolsep}{1.54pt}
\renewcommand{\arraystretch}{1.2}
\begin{tabular*}{\linewidth}{@{\extracolsep{\fill}}cccc|cccc}

\multicolumn{4}{c}{\textbf{Velocity function (Fig. \ref{fig:CVF_morph_SF})}}                                                              & \multicolumn{4}{c}{\textbf{Halo mass function (Fig. \ref{fig:HMF})}}                                                             \\
\hline
\hline

\begin{tabular}[c]{@{}c@{}}Bin centre\\ $\log_{10}(v_{\rm{circ}}/\rm{km\ s^{-1}})$\end{tabular} & \begin{tabular}[c]{@{}c@{}} Number density $\phi$ \\ $\mathrm{Mpc^{-3}dex^{-1} }h_{70}^{3}$ \end{tabular} & $\sigma_{\log_{10}(v_{\rm{circ}})}$  & $\sigma_{\phi}\times 10^{-3}$

& \begin{tabular}[c]{@{}c@{}}Bin centre\\ $\log_{10}(M_{\rm{200}}/\rm{\rm{M_{\odot}}}h_{70}^{-1})$ \end{tabular} & \begin{tabular}[c]{@{}c@{}} Number density $\phi$ \\ $\mathrm{Mpc^{-3}dex^{-1} }h_{70}^{3}$ \end{tabular} & $\sigma_{\log_{10}(M_{\rm{200}})}$ &  $\sigma_{\phi} \times 10^{-4}$ \\
\midrule

2.04                                                            &  $3.2 \times 10^{-2}$   &  (+0.02 -0.03)                &  (+2.9 -2.9)         & 11.53                                                                &  $1.19\times10^{-2}$   &     (+0.07 -0.13)                            &  (+6.1 -6.3)        \\
2.08                                                            &  $3.4\times10^{-2}$   &  (+0.02 -0.02)                &    (+2.2 -2.1)      &   11.71                                                             & $1.11\times10^{-2}$     &        (+0.09 -0.11)          &   (+5.1 -5.1)         \\
2.12                                                            &  $3.4\times10^{-2}$   & (+0.02 -0.02)                  &   (+2.0 -1.9)       &    11.90                                                            & $7.6\times10^{-3}$     &     (+0.09 -0.10)             & (+3.4 -3.5)             \\
2.17                                                            &   $3.0\times10^{-2}$  &  (+0.02 -0.03)                 &    (+2.1 -2.0)      &  12.10                                                                                                                     & $6.2\times10^{-3}$    &        (+0.10 -0.09)            &  (+3.0 -3.0)         \\
2.21                                                            &   $2.0\times10^{-2}$  &  (+0.02 -0.02)                 &     (+1.1 -1.1)     & 12.30                                                               & $3.6\times10^{-3}$    &       (+0.10 -0.09)           & (+1.8 -1.9)         \\
2.26                                                            &   $2.1\times10^{-2}$  &  (+0.02 -0.02)                 &    (+1.0 -1.1)      &   12.50                                                             & $1.9\times10^{-3}$      &       (+0.09 -0.10)            & (+1.1 -1.2)          \\ 2.30
                                                            & $2.2\times10^{-2}$     & (+0.02 -0.02)                   &     (+1.1 -1.0)     &      12.71                                                          & $9.2\times10^{-4}$    & (+0.10 -0.11)                  & (+6.0 -6.0)e-1          \\
2.35                                                            &  $1.96\times10^{-2}$     & (+0.02 -0.03)                  &    (+9.3 -9.4)e-1       &    12.89                                                            &   $3.7\times10^{-4}$  &      (+0.11 -0.09)            & (+2.9 -2.9)e-1           \\
2.39                                                            &  $1.01\times10^{-2}$    & (+0.02 -0.03)                 &      (+4.9 -5.0)e-1     &  13.07                                                               &  $8.1\times10^{-5}$    &          (+0.13 -0.07)         &       (+8.4 -8.0)e-2   \\
2.43                                                            & $4.5\times10^{-3}$     & (+0.02 -0.02)                 &     (+2.8 -3.0)e-1      &  13.28                                                              & $1.4\times10^{-5}$      &         (+0.12 -0.08)          &       (+3.2 -3.0)e-3    \\ 2.48
                                                               & $9.4\times10^{-4}$      &(+0.03 -0.03)                  &      (+9.0 -8.6)e-2     &13.48                                                                &$2.2\times10^{-6}$     & (+0.12 -0.08)                  & (+9.4 -9.9)e-3      \\ 2.51
                                                               & $5.6\times10^{-5}$      & (+0.04 -0.006)                  &     (+9.4 -9.3)e-3     &                                                                &     &                  &   \\

\hline
\hline
\end{tabular*}
\end{table*}

The disagreement between our HMF and that of \cite{li_halo_2019} in the range $10^{11.5}\leqslant M_{200}/\rm{M_{\odot}} \leqslant 10^{12.6}$ is due to a lack of completeness in the HIPASS sample (as it does not include massive quenched ET galaxies), also seen when comparing the respective CVFs (Fig. \ref{fig:CVF}). The contribution from SF galaxies to our HMF shows a higher number density still than the \cite{li_halo_2019} HMF above $10^{11.7}$. This difference is due to the reduced sensitivity of the HIPASS survey in the high mass regime compared to e.g. ALFALFA (as discusses in Section \ref{sec: Circular velocity function}). 
The lower number densities predicted by the MaNGA HMF compared to that of \cite{driver_empirical_2022} above $ 10^{12.7}\rm{M_{\odot}}$ is due to our $M_{200}$ calculation reflecting the halos of individual galaxies (or the contribution of an individual galaxy to the halo of the group to which it belongs). In contrast, the \cite{driver_empirical_2022} measurements reflect the number density of group/cluster halos with a respective $M_{200}$. 

Our HMF computation covers $\sim 1 $ dex ($10^{11.7}-10^{12.7}\rm{M_{\odot}}$) of previously un-probed (or probed with a lack of completeness) halo masses. The MaNGA HMF in this range is remarkably well in agreement with the extrapolation of the best-fitting MRP function from \cite{driver_empirical_2022}, and with the $\Lambda$CDM-based HMF of \cite{murray_thehalomod_2021}. However, these HMFs are slightly over-estimating (by a factor of 1.8 to 2.4) the number density of halos compared to \cite{li_halo_2019} below the lowest MaNGA halo mass threshold. This discrepancy is contributed to by the fact that $M_{200}$ masses from \cite{li_halo_2019} only include dark matter. The addition of baryons, while systematically shifting the \cite{li_halo_2019} HMF to higher $M_{200}$, would only produce a 1-5 per cent increase (Section \ref{sec: SMHM rel}). In the next section, we discuss the contribution of different morphological and  star forming galaxy classes to the mass density in the nearby Universe.  

For the remainder of this paper, we switch to a Planck 2018 cosmology (\citealt{aghanim_planck_2020}) with $h_{\rm{67}}$ = $H_{\rm{0}}/(67.37$ $\rm{km\ s^{-1}\ Mpc^{-1}})$, to facilitate comparison with the latest estimate of the present-day matter density.

\subsection{Halo mass function - contribution from different galaxy classes to the density of matter $\rho_{\rm{M}}$}
\label{sec:HMF_separate}

A powerful prediction of the HMF is its first moment, i.e. the integration of the HMF over halo masses. This integral is equivalent to the matter density ($\rho_{\rm{M}}$) of the mass interval within which it is performed. The integral of the MRP function (equation \ref{eq:MRP}) between two halo masses $M_{\rm{low}}$ and $M_{\rm{high}}$ can be calculated analytically as:

\begin{equation}
    \rho_{\rm{M}} = \phi_{\star} M_{\star} \Bigg[ \Gamma\Bigg( \frac{\alpha+2}{\beta}, \bigg( \frac{M_{\rm{high}}}{M_{*}} \bigg)^{\beta}  \Bigg) -  \Gamma \Bigg( \frac{\alpha+2}{\beta},    \bigg( \frac{M_{\rm{low}}}{M_{*}} \bigg)^{\beta}     \Bigg)  \Bigg].
    \label{eq:integral}
\end{equation}

\noindent In the above equation, $\phi_{\star}$, $M_{\star}$, $\alpha$ and $\beta$ are the parameters of the MRP function in equation \ref{eq:MRP}, and $\Gamma$ is the lower-incomplete Gamma function.

\begin{figure}
	\centering
	\includegraphics[width=\columnwidth]{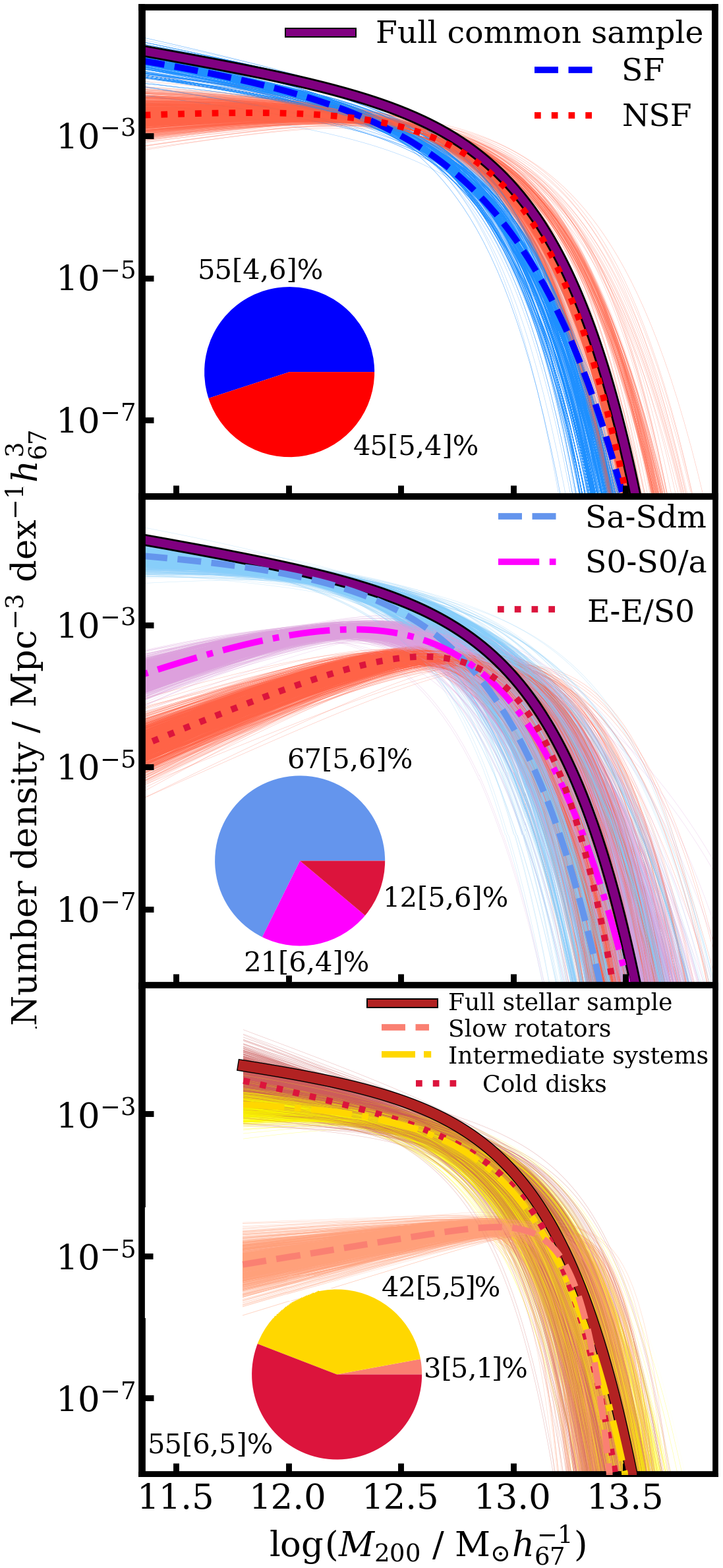}
    \caption{The best-fitting MRP function to the HMF of the MaNGA \textcolor{violet}{\textbf{common}} sample (upper two panels) and \textcolor{Maroon}{\textbf{stellar sample}} (bottom panel), split between contributions from: SF and NSF galaxies (\textbf{top}); E-E/S0, S0-S0/a and Sa-Sdm morphologies (\textbf{middle}) and kinematic classes (\textbf{bottom}). The thin curves are the result of scattering the HMF data points (computed using \texttt{dftools}) about their uncertainties in $M_{200}$ (assuming a lognormal distribution), and re-performing the fit (repeated 1000 times). The pie charts show the contribution of each sub-sample (matched in colour to the curves) to the matter density $\rho_{\rm{M}}$ (equation \ref{eq:integral}) for $\log_{10}(M_{200}/ \mathrm{M_{\odot}} h_{67}^{-1})$ between 11.35-13.50 (upper two panels) and 11.80-13.50 (bottom panel). The results are consistent within 0.5 per cent if we instead integrate to infinity. The numbers in square brackets reflect the upper (first) and lower (second) uncertainties on the percentages, calculated as the spread between the $16^{\rm{th}}$ and $84^{\rm{th}}$ percentiles of the $\rho_{\rm{M}}$ distributions for the 1000 scattered MRP fits.  } 
    \label{fig:HMF_split_classes}
\end{figure}

Fig. \ref{fig:HMF_split_classes} shows the HMF for our common kinematic sample, split between contributions from SF and NSF galaxies (top), and between contributions from three morphological classes representative of elliptical (red, $\rm{T-type} \leqslant -1.5$), lenticular (pink, $-1.5< \rm{T-type} \leqslant 0.5$) and spiral (blue, $\rm{T-type} > 0.5$) galaxies. We integrate (equation \ref{eq:integral}) the best-fitting MRP functions for each sub-sample within the halo mass range probes by our data ($10^{11.35}\rm{M_{\odot}}-10^{13.50}\rm{M_{\odot}}$). For both sample splits, the sum of the integrals from sub-samples adds up to a fraction of the integral for the full sample fit which is consistent with 1, without imposing any priors (1.005 and 1.030 for the SF/NSF and morphological separations, respectively).

\begin{table}
\centering
\caption{ Mass density ($\rho_{\rm{M,sample}}$) associated with the sub-samples shown in Fig. \ref{fig:HMF_split_classes}, in the range $10^{11.35}\leqslant M_{200}/ M_{\odot}\leqslant 10^{13.5}$ (full sample) and $10^{11.80}\leqslant M_{200}/ M_{\odot}\leqslant 10^{13.5}$ (stellar sample). The corresponding contribution to the density parameter is also displayed ($\Omega_{\rm{M,sample}}=\rho_{\rm{M,sample}}/\rho_{\rm{crit}}$). The numbers in square brackets are the upper (first) and lower (second) uncertainties, calculated as the $16^{\rm{th}}$ and $84^{\rm{th}}$ percentiles of the $\rho_{\rm{M,sample}}$ and $\Omega_{\rm{M,sample}}$ distributions from the 1000 fits of the scattered HMFs in Fig. \ref{fig:HMF_split_classes}. }
\label{tab:rho_m_split}

\begin{tabular}{@{\extracolsep{\fill}}ccc}
Sample & $\rho_{\rm{M,sample}}\times 10^{9} \mathrm{M_{\odot}}\ \rm{Mpc^{-3}}  $ $h_{67}^{2}$ & $\Omega_{\rm{M,sample}}$  \\
\midrule
\midrule
Full           &  9.3$[0.2,0.3]$              & 0.065$[0.001,0.002]$         \\
\midrule
SF          &   5.1$[0.4,0.6]$               & 0.036$[0.002,0.004]$            \\
NSF              &   4.2$[0.5,0.4]$             &   0.030$[0.002,0.003]$         \\
\midrule
Sa-Sdm         &    6.2$[0.5,0.5]$            & 0.044$[0.003,0.004]$           \\
S0                  &  2.0$[0.6,0.4]$              & 0.014$[0.004,0.003]$         \\
E-E/S0                  &  1.1$[0.5,0.5]$              & 0.008$[0.003,0.004]$         \\
\midrule
\midrule
Stellar                  &  4.9$[0.3,0.2]$              & 0.035$[0.002,0.003]$  \\

Slow rot.                  &  0.15$[0.3,0.02]$              & 0.001$[0.003,0.0003]$ \\

Int. syst.                  &  2.1$[0.2,0.3]$              & 0.015$[0.001,0002]$ \\ 

Cold discs                  &  2.7$[0.3,0.2]$              & 0.019$[0.002,0.002]$  \\
\midrule
\midrule
\end{tabular}
\end{table}

SF galaxies dominate the number density of galaxies up to $10^{12.3}\rm{M_{\odot}}$, after which NSF objects become dominant. When splitting by morphology, spirals dominate the number density of nearby-Universe galaxies, followed by lenticulars and ellipticals, up to $ 10^{12.8}\rm{M_{\odot}}$, when the contributions of the three classes are equal. At higher $M_{200}$, the trend in the number density is reversed.

The resulting mass densities $\rho_{\rm{M,sample}}$ and fractional matter densities $\Omega_{\rm{M,sample}}=\rho_{\rm{M,sample}}/\rho_{\rm{crit}}$ (where $\rho_{\rm{crit}}$ is the critical density of the Universe at the present day) for each sub-sample in Fig. \ref{fig:HMF_split_classes} are displayed in Table \ref{tab:rho_m_split}. As discussed in Appendix \ref{sec:Appendix_errors}, the matter density and fractional density parameters in Table \ref{tab:rho_m_split} change by up to 9 per cent when systematic uncertainties in halo mass are taken into account. SF and NSF galaxies contribute $55 \substack{+4 \\ -6}$ and $45 \substack{+5 \\ -4}$ per cent of the matter density of galaxies with halo masses between $10^{11.35}-10^{13.50}\rm{M_{\odot}}$. The fractional contribution from elliptical, lenticular and spiral morphologies are $12 \substack{+5 \\ -6}$, $21 \substack{+6 \\ -4}$ and $67 \substack{+5 \\ -6}$ per cent, respectively.

A similar analysis in terms of stellar mass for galaxies down to $M_{\star} = 10^{6}\ \rm{M_{\odot}}$ by \cite{driver_galaxy_2022} using the GAMA galaxy survey found that galaxies with a disc component (including systems with a compact or diffuse bulge) contribute 67 per cent to the stellar mass budget, with early-type systems contributing 32 per cent. 
Our results for halo mass are in relative agreement with those of \cite{driver_galaxy_2022}, considering the almost linear SMHM relation of Fig. \ref{fig:stellar_mass_halo_mass} (partly by construction, given our assumption of the $R_{\mathrm{e}}-R_{200}$ relation of \citealt{kravtsov_size-virial_2013} for ETs, which in turn assumes the abundance matching ansatz), with the difference that our sample is only representative of the nearby-Universe population above $M_{\star} =10^{9}\ \mathrm{M_{\odot}}$. Lenticular (S0/S0a) systems in our sample are more typical of the ET population, as defined in \cite{driver_galaxy_2022}. A split between LTs and ETs at T-type = 0 results in a 67-33 per cent split of the mass density between the two morphological classes. Extremely compact and highly asymmetrical systems included in the analysis of \cite{driver_galaxy_2022} are not present in our sample; however, those classes have been found to contribute <1 per cent to the number density of galaxies at all stellar masses.

The SF and NSF populations reach equal number densities at $M_{\rm{200}}\approx 10^{12.3}\ \rm{M_{\odot}}$, which corresponds to a stellar mass of $\approx 10^{10.7}\ \rm{M_{\odot}}$ based on the SMHM relation for our data (Fig. \ref{fig:stellar_mass_halo_mass}). The same equality between the three morphological categories in Fig. \ref{fig:HMF_split_classes} is reached for $M_{\rm{200}}\approx 10^{12.8}\ \rm{M_{\odot}}$, corresponding to $M_{\star}\approx 10^{11.2} \ \rm{M_{\odot}}$. The dominance of spiral galaxies below this threshold supports the findings of \cite{rigamonti_bang-manga_2024}, who reported that for $M_{\star}\leqslant 10^{11}\ \rm{M_{\odot}}$ there is little difference in terms of kinematic morphology between active and passive galaxies.

In the bottom panel of Fig. \ref{fig:HMF_split_classes} we show the HMF for our stellar kinematic sample. The minimum stellar mass of this sample, $10^{10.1}\ \mathrm{M_{\odot}}$ (Table \ref{tab:samples_completeness}), corresponds to a halo mass of $\approx 10^{11.8}\ \mathrm{M_{\odot}}$. We elect to exclude the galaxies from our stellar sample with $M_{\rm{200}} < 10^{11.8}\ \mathrm{M_{\odot}}$ to avoid a lack of completeness in the low mass end. The sample in the bottom panel of Fig. \ref{fig:HMF_split_classes} includes 1673 galaxies and is statistically complete for $M_{\star} \geqslant 10^{10.2} \ \mathrm{M_{\odot}}$. The number density of galaxies in the probed halo mass range is dominated by cold discs below $\approx 10^{13.2} \ \mathrm{M_{\odot}}$, above which all kinematic morphologies contribute equal amounts, within uncertainties. We find that slow rotators contribute only 3$\substack{+5 \\ -1}$ per cent of the matter density for galaxies with $M_{\rm{200}}\geqslant10^{11.8}\ \mathrm{M_{\odot}}$ ($M_{\star} \geqslant 10^{10.2}\ \mathrm{M_{\odot}}$) in the nearby Universe. The majority of the matter density is accounted for by cold discs (55$\substack{+6 \\ -5}$ per cent) and intermediate systems (+42$\substack{+5 \\ -5}$ per cent), showing that processes causing the creation of slow rotators (mergers, secular heating; see e.g. \citealt{cortese_physical_2022}, \citealt{rutherford_sami_2024}, \citealt{croom_sami_2024}) have a relatively low occurrence rate for galaxies in this mass range.

In summary, our results show that SF and NSF galaxies above $M_{\star} = 10^{9.2}\ \rm{M_{\odot}} (M_{200} = 10^{11.35}\ \rm{M_{\odot}})$ contribute almost equal amounts to the density of matter $\rho_{\rm{M}}$ in the nearby Universe. Spiral morphologies (Sa-Sdm) are contributing 67 per cent to $\rho_{\rm{M}}$, while the contribution of early-types (lenticulars and ellipticals, E-S0/a) is 33 per cent. Kinematically cold discs and intermediate systems dominate the mass density in the nearby Universe (55 and 42 per cent, respectively), for galaxies with stellar masses above $10^{10.2}\ \mathrm{M_{\odot}}$. Slow rotators in this mass range only account for 3 per cent of the matter density.
In the next section, we combine the MaNGA HMF derived in this work with those of \cite{li_halo_2019} and \cite{driver_empirical_2022} to constrain the HMF across 5 orders of magnitude in halo mass.

\begin{figure*}
	\centering
	\includegraphics[width=\linewidth]{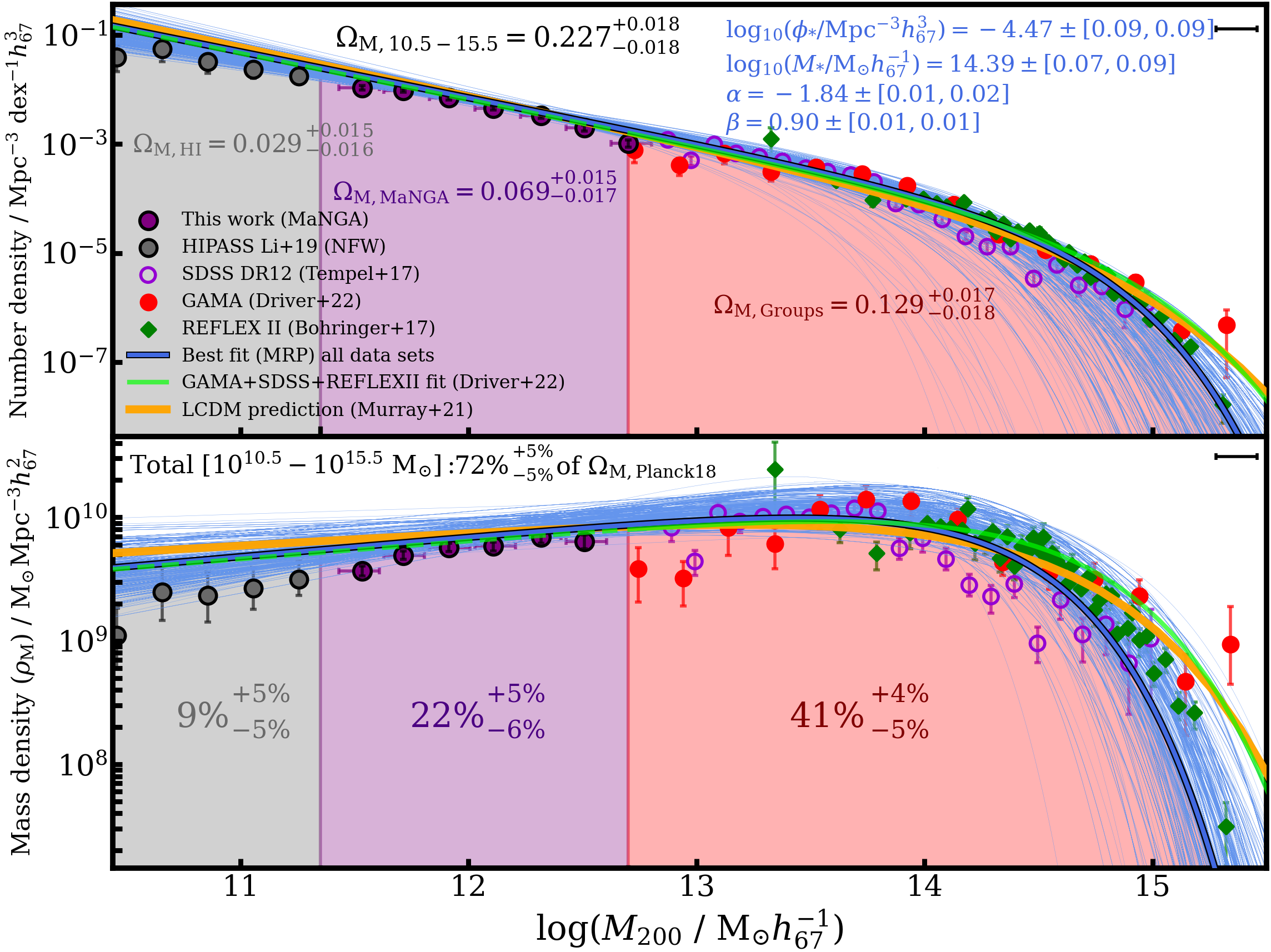}
    \caption{ \textbf{Top:} The HMF combining measurements from HIPASS (\citealt{li_halo_2019}, gray circles) for $10.50 \leqslant \log_{10}(M_{200}/\rm{M_{\odot}}) \leqslant 11.35$, this work (MaNGA, purple circles) for $11.35 \leqslant \log_{10}(M_{200}/\rm{M_{\odot}}) \leqslant 12.70$ and the measurements from group/cluster masses compiled by \protect\cite{driver_empirical_2022} for $12.7 \leqslant \log_{10}(M_{200}/\rm{M_{\odot}}) \leqslant 15.5$. The thin blue curves are the result of randomly scattering the HMF measurements about their uncertainties in $\log_{10}(M_{200})$ 1000 times (assuming a lognormal distribution), and re-fitting the combined data sets. The black error bar in the upper right corner is the assumed $\log_{10}(M_{200})$ uncertainty in the data points not derived in this work. The solid blue curve is the best-fitting MRP function through the combined data sets (the median of the distributions from the 1000 individual fittings), with the best-fitting parameters displayed. The numbers in square brackets show the upper (first) and lower (second) uncertainties in the best-fitting parameters. The green curve is the best-fitting HMF from \protect\cite{driver_empirical_2022}, which incorporates the $\Omega_{\rm{M,Planck18}}$ value as a prior (dashed below the lower $M_{200}$ limit of the fitted data). The orange curve shows the $\Lambda$CDM HMF prediction of \protect\cite{murray_thehalomod_2021}. The coloured areas reflect the extent covered by each data set. The contribution to the density parameter $\Omega_{\rm{M,Planck18}}$ at the present day from each $M_{200}$ interval is displayed. The sum of these contributions is displayed at the top. 
    \textbf{Bottom:} Similar to the top, but with the vertical axis now showing the mass density. The contribution from the ranges in $\log_{10}(M_{200})$ covered by each data set to the fractional density of matter reported by the Planck collaboration (\protect\citealt{aghanim_planck_2020}, $\Omega_{\rm{M,Planck18}}$ = 0.3147) is displayed. For the full range between $10.5 \leqslant \log_{10}(M_{200}/\rm{M_{\odot}}) \leqslant 15.5$, we recover $72 \substack{+3 \\ -4}$ per cent of the matter density predicted by the Planck collaboration. To facilitate a comparison with results of the Planck collaboration, we scale parameters in this plot with respect to $h_{67}=H_{0}/(67.4\ \rm{km\ s^{-1}\ Mpc^{-1})}$  } 
    \label{fig:HMF_Integral}
\end{figure*}

\begin{figure}
	\centering
	\includegraphics[width=\columnwidth]{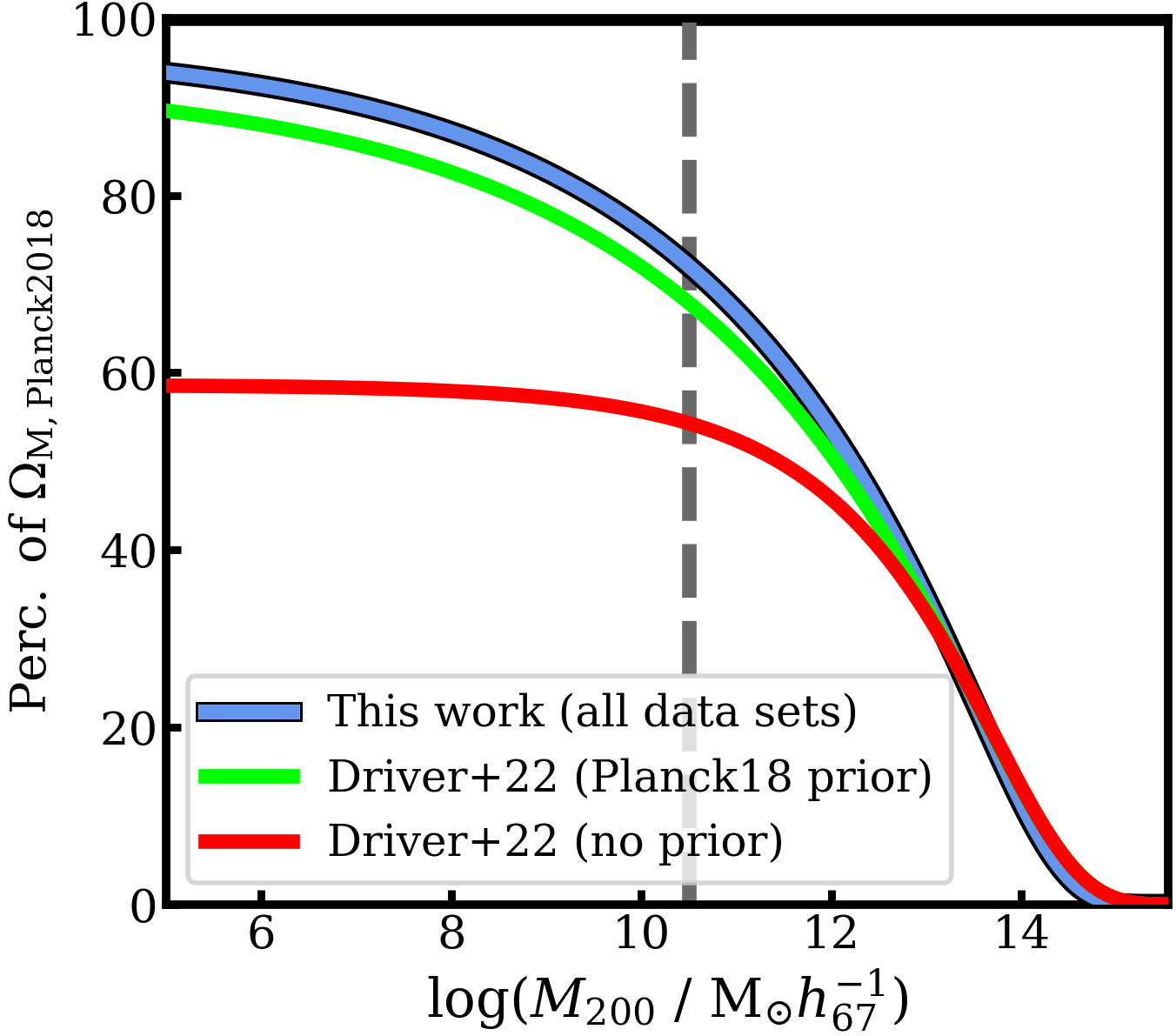}
    \caption{ The percentage of the matter density parameter predicted by \citealt{aghanim_planck_2020} ($\Omega_{\mathrm{M,Planck18}}=0.3147 \pm 0.0074$) that the displayed HMF fits reach if integrated down to the halo mass on the horizontal axis. The blue and green curves represent the best-fitting HMFs with the same colour in Fig. \ref{fig:HMF_Integral}. The red curve is the best-fitting HMF from \protect\cite{driver_empirical_2022} without any priors. The gray dashed line is placed at the lowest $M_{200}$ limit of the HIPASS HMF from \protect\cite{li_halo_2019}.   } 
    \label{fig:integral}
\end{figure}

\subsection{Halo mass function - constraints across 5 dex in $\log_{10}(M_{200})$ and contribution to $\Omega_{\rm{M}}$}
\label{sec:HMF_Omega}

We now leverage the information afforded by the MaNGA HMF in extending the group and cluster HMF of \cite{driver_empirical_2022}, which combines data from the GAMA, SDSS and REFLEX II surveys, to the range $10^{11.35} \rm{M_{\odot}} \leqslant M_{200} \leqslant 10^{12.70} \rm{M_{\odot}}$. Additionally, the inclusion of the HIPASS HMF from \cite{li_halo_2019} below the lower limit of the MaNGA one allows us to constrain the HMF in the range $10^{10.5} \rm{M_{\odot}} \leqslant M_{200} \leqslant 10^{15.5} \rm{M_{\odot}}$. 

In Fig. \ref{fig:HMF_Integral} we show the combined HMF for the five data sets. We consider measurements from GAMA, SDSS and REFLEX II for $M_{200}\geqslant10^{12.7}\ \mathrm{M_{\odot}}$, corresponding to the lower limit above which the GAMA group catalogue is complete (\citealt{driver_empirical_2022}). The median halo occupation density at this mass threshold predicted by \textit{The Next Generation Illustris} (IllustrisTNG, \citealt{marinacci_first_2018, naiman_first_2018, nelson_first_2018,springel_first_2018,pillepich_first_2018}) suite of cosmological, hydrodynamical simulations of galaxy evolution is $\sim 2.4$ ($\sim2.0$ at $10^{12.6}\ \mathrm{M_{\odot}}$ and $\sim1.5$ at $10^{12.3} \ \mathrm{M_{\odot}}$, \citealt{bose_revealing_2019}), consistent with the transition from groups to single-occupation halos. Between $10^{11.35} - 10^{12.70} \rm{M_{\odot}}$ we only consider the HMF data derived in this work from the MaNGA survey, due to the lack of completeness in HIPASS data within this mass range. 

The combined HMF data in shown in Table \ref{tab:functions_data}. To model the HMF with an MRP function, we scatter the HMF data points computed in this work (purple) about their uncertainties in $\log_{10}(M_{200}/\mathrm{M_{\odot}})$ (assuming a lognormal distribution) and fit an MRP using least-squares regression at each iteration. The uncertainties in halo mass reflect the width of the bins selected for generating the HMF, which in turn were chosen such that half the bin width ($\sim$0.1 dex) is equal to the average uncertainty in $\log(M_{200}/\mathrm{M_{\odot}})$ for our sample. No such error estimates are available for the literature data sets (GAMA, SDSS, REFLEX II) considered in this analysis. For the purpose of the scattering procedure, the uncertainties in $\log_{10}(M_{200}/\mathrm{M_{\odot}})$ associated with these literature data sets are assumed to be equal to the median  $\log_{10}(M_{200}/\mathrm{M_{\odot}})$ uncertainty corresponding to the HMF data points computed in this work (the black error bar in the upper right corner of Fig. \ref{fig:HMF_Integral}). The scattering process about uncertainties in $\log_{10}(M_{200}/\mathrm{M_{\odot}})$ is repeated 1000 times (thin blue lines), and the best-fitting MRP is defined by the median of the parameter distributions from the 1000 iterations (thick blue line, with best-fitting parameters displayed). If we instead perform this process by scattering only the MaNGA HMF data points about uncertainties in $\log(M_{200}/\mathrm{M_{\odot}})$, the best-fitting joint HMF parameters remain unchanged, while their associated uncertainties decrease by 1.5 - 3.7 times. 
The best-fitting parameters of the joint HMF in Fig. \ref{fig:HMF_Integral} change by up to 2 per cent if systematic uncertainties in halo mass are considered, as detailed in Appendix \ref{sec:Appendix_errors} (Table \ref{tab:systematic_effect}).

We integrate under the best-fitting MRP function to determine the matter density $\rho_{\rm{M}}$ corresponding to halos within a given range in $M_{200}$ (equation \ref{eq:integral}). The contribution to the fractional matter density in the nearby Universe, $\Omega_{\rm{M}}$, from halos in this range is calculated as $\rho_{\rm{M}}/\rho_{\rm{crit}}$. $\Omega_{\rm{M}}$ contributions from the $M_{200}$ ranges probed by each data set are highlighted in Fig. \ref{fig:HMF_Integral}. Within the halo mass range probed, we recover a fractional matter density $\Omega_{\rm{M,10.5-15.5}} = 0.227 \pm 0.018$ ($\rho_{\rm{M}}=3.2 \pm 0.2   \times 10^{10} \rm{M_{\odot}\ Mpc^{-3}}  $ $h_{67}^{2}$). As described in Appendix \ref{sec:Appendix_errors}, the fractional matter density $\Omega_{\rm{M,10.5-15.5}}$ changes by up to 12 per cent when systematic uncertainties in $M_{200}$ are considered.

The best-fitting HMF fit derived in this work agrees remarkably well with that of \cite{driver_empirical_2022} for groups and clusters, which incorporates the $\Omega_{\rm{M}}$ value of \cite{aghanim_planck_2020} as a prior (green curve in Fig. \ref{fig:HMF_Integral}). 
The largest difference is in the exponential softening parameter $\beta$ (0.90$ \substack{+0.01 \\ -0.01}$ in this work compared to $0.77\pm 0.11$ in \citealt{driver_empirical_2022}; the other three best-fitting parameters agree within 0.7$\sigma$). We also find a good agreement with the $\Lambda$CDM-based HMF of \cite{murray_thehalomod_2021}. All of the best-fitting parameters are consistent within $1 \sigma$, except the high-mass cut-off $\beta$ which is consistent within  $1.7 \sigma$. This difference translates to a lower galaxy number density in the high-$M_{200}$ end compared to the $\Lambda$CDM prediction of \cite{murray_thehalomod_2021}, placing constraints on the contribution of mergers in forming the most massive halos in the nearby Universe. The difference between the high-mass cut-off of our HMF and that of the HMF inferred by \cite{murray_thehalomod_2021} also corresponds to a ratio between $\Omega_{\mathrm{M}}$ (integrated to zero mass) values (determined in this work to that predicted by \citealt{murray_thehalomod_2021}) of 0.89. Taking into account the constraints on the Sigma-8 tension $S_8 \equiv \sigma_{8} (\Omega_{\mathrm{M}}/0.3)^{0.5}$ (where $\sigma_8$ is the root mean square fluctuation of the mass distribution in the Universe on scales of $8\ h^{-1}\ \mathrm{Mpc}$) determined by \cite{aghanim_planck_2020}, the ratio of $\sigma_8$ values (inferred in this work to that predicted by \citealt{murray_thehalomod_2021}) is 1.06. Both of these ratios are consistent with 1 when taking into account either the statistical or the systematic (Appendix \ref{sec:Appendix_errors}) uncertainties associated with our measurements. 

With the constraints of the MaNGA HMF derived in this work, and those of the HIPASS HMF from \cite{li_halo_2019}, integrating the best-fitting MRP function from Fig. \ref{fig:HMF_Integral} to zero mass results in a value of the density parameter $\Omega_{\rm{M}}=0.306 \substack{+0.018 \\ -0.019}$. This value is consistent with that of \cite{aghanim_planck_2020} ($\Omega_{\rm{M,Planck18}} = 0.3147 \pm 0.0074$) within uncertainties, without imposing any priors in the minimisation procedure. The $\Omega_{\rm{M}}$ value changes by up to 8 per cent when systematic uncertainties in halo masses for the MaNGA sample are considered, as detailed in Appendix \ref{sec:Appendix_errors} (Table \ref{tab:systematic_effect}). The integral of our best-fitting joint HMF down to a given $M_{200}$ is illustrated in Fig. \ref{fig:integral}, as a percentage of the matter density parameter from \cite{aghanim_planck_2020}. 94 per cent of the matter density in the nearby Universe is accounted for by the best-fitting HMF derived in this work when integrating down to a halo mass of $10^{5.0}\ \rm{M_{\odot}}$. This result is similar to that obtained when integrating  the best-fitting HMF from \cite{driver_empirical_2022} (reaching 90 per cent of $\Omega_{\rm{M,Planck18}}$ down to $10^{5.0} \ \rm{M_{\odot}}$), which includes the prior that the difference with respect to $\Omega_{\rm{M,Planck18}}$ is minimised in the fitting procedure (contrary to the HMF derived in this work, where no priors are included in the fitting). However, our result differs greatly from the \cite{driver_empirical_2022} HMF with no priors, which only reaches $\approx60$ per cent of this value. The best-fitting MRP functions to the CVF and HMF derived in this work are shown in Table \ref{tab:vf_hmf_bf}.

In Fig. \ref{fig:HMF_Integral} (bottom) we show the mass density ($\rho_{\rm{M}}$, number density multiplied by corresponding $M_{200}$) as a function of halo mass. The dominant contribution to the matter density comes from group and cluster halos ($M_{200} \geqslant 10^{12.5}\rm{M_{\odot}} $), with $\Omega_{\rm{M,Groups}} = 0.129 \substack{+0.017 \\ -0.018}$, equivalent to $41 \substack{+4 \\ -5} $ of the matter density parameter derived by \cite{aghanim_planck_2020}. This value is in agreement with the result of \cite{driver_empirical_2022} for $M_{200}\geqslant10^{12.7}\rm{M_{\odot}}$ ($0.128 \pm 0.016$).

Dark matter halos occupied by a single galaxy probed by the MaNGA (this work) and HIPASS (\citealt{li_halo_2019}) surveys ($10^{10.5}\rm{M_{\odot}}\leqslant M_{200} \leqslant 10^{12.7} \rm{M_{\odot}}$), albeit more numerous than large group or cluster halos, contribute by only 31$\substack{+5 \\ -6}$ per cent to the present-day matter density ($\Omega_{\rm{MaNGA+HIPASS}} = 0.098 \substack{+0.016 \\ -0.018}$ ). 
Our combined HMF for the halo mass range shown in Fig. \ref{fig:HMF_Integral} encompasses $72\substack{+5 \\ -5}$ per cent of the expected matter density in the present-day Universe (\citealt{aghanim_planck_2020}).

We perform two final checks to ensure the robustness of our HMF computation. Pertaining to the caveats of the \cite{li_halo_2019} HMF, where $M_{200}$ is an estimate of dark matter mass and does not include baryons (Section \ref{sec: SMHM rel}), we fit the HMF in Fig. \ref{fig:HMF_Integral} excluding the HIPASS data. We obtain results consistent within uncertainties, as detailed in Table \ref{tab:vf_hmf_bf} (bottom row). Considering that our HMF measurement from the MaNGA sample are representative of single-occupation halos rather than groups, we considered the approach of adding the MaNGA number densities for halos with $M_{200}\geqslant10^{12.7}$ in Fig. \ref{fig:HMF} to the best-fitting HMF fit of \cite{driver_empirical_2022} (without any priors) and including these data points in the joint HMF fit. The resulting MRP fit agrees with the blue line of  Fig. \ref{fig:HMF_Integral} remarkably well (best-fitting parameters are consistent within 0.4$\sigma$).

In summary, we provide a computation of the HMF spanning 5 orders of magnitude in $M_{200}$ ($10^{10.5} \rm{M_{\odot}} \leqslant M_{200} \leqslant 10^{15.5} \rm{M_{\odot}}$). The HMF is well modelled by an MRP function (Fig. \ref{fig:HMF_Integral}), and within the $M_{200}$ range probed we resolve 72 per cent of the mass density predicted by \cite{aghanim_planck_2020} for the present day Universe, corresponding to $\Omega_{\rm{M,10.5-15.5}} = 0.227 \pm 0.018$. The contribution to the mass density from single-occupation halos below the lower limit of \cite{driver_empirical_2022} is equivalent to 31 per cent of the Universe's matter density, while groups and clusters contribute 41 per cent. Without imposing any priors, the HMF derived in this work integrates to a total matter density consistent with the measurement of \cite{aghanim_planck_2020}, within uncertainties, and agrees with the $\Lambda$CDM prediction of \cite{murray_thehalomod_2021}.

\begin{table*}
\centering
\caption{Compilation of the best-fitting parameters of the circular velocity and halo mass functions derived in this paper, from a Murray-Robotham-Power (MRP) fit (equation \ref{eq:MRP}). The best-fitting parameters of the halo mass function for the MaNGA sample are expressed in terms of $h_{\mathrm{x}}=h_{70}=H_{0}/(70\ \rm{km\ s^{-1}\ Mpc^{-1}}$). For the other three HMF fits: (\textbf{MaNGA+HIPASS+GAMA+SDSS+REFLEX II}),  (\textbf{MaNGA+HIPASS}), and (\textbf{MaNGA+GAMA+SDSS+REFLEX II}), best-fitting parameters are expressed in terms of $h_{\mathrm{x}}=h_{67}=H_{0}/(67.37\ \rm{km\ s^{-1}\ Mpc^{-1}})$ to facilitate comparison with the latest cosmological parameters reported by the Planck collaboration (\citealt{aghanim_planck_2020}).    }
\label{tab:vf_hmf_bf}
\setlength{\tabcolsep}{6.8pt}
\renewcommand{\arraystretch}{1.4}
\begin{tabular*}{\linewidth}{@{\extracolsep{\fill}}cccccc}

\hline
\hline
\multicolumn{6}{c}{\textbf{Velocity function}}                                                             
\\
\midrule
\midrule
Sample                                                                                                & \multicolumn{1}{l}{Appears in Fig.} & $\log_{10}(\phi_{*}/\mathrm{Mpc^{-3} }h_{70}^{3})$        & $\log_{10}(M_{*}/\mathrm{M_{\odot} }h_{70}^{-1})$     &   $\alpha$           & $\beta$           \\

\textbf{\textcolor{violet}{MaNGA}  }                                                                                               & \ref{fig:CVF_morph_SF}                                   & -2.99 $\substack{+0.08 \\ -0.07}$  & 2.420 $\substack{+0.005 \\ -0.007}$ & -1.7 $\substack{+0.1 \\ -0.1}$  & 7.9 $\substack{+0.5 \\ -0.7}$ \\
\midrule
\midrule
\multicolumn{6}{c}{ \textbf{Halo mass function} }\\ 
\midrule
\midrule
Sample                                                                                                & \multicolumn{1}{l}{Appears in Fig.} & $\log_{10}(\phi_{*}/\mathrm{Mpc^{-3} }h_{\mathrm{x}}^{3})$        & $\log_{10}(M_{*}/\mathrm{M_{\odot} }h_{\mathrm{x}}^{-1})$     &   $\alpha$           & $\beta$            \\

\textbf{\textcolor{violet}{MaNGA}}                                                                                                 & \ref{fig:HMF}, \ref{fig:HMF_split_classes}                                  & -2.97 $\substack{+0.09 \\ -0.09}$   & 12.69 $\substack{+0.06 \\ -0.09}$  & -1.54 $\substack{+0.03 \\ -0.05}$ & 1.26 $\substack{+0.04 \\ -0.02}$ \\
\midrule
\multicolumn{1}{l}{\begin{tabular}[c]{@{}c@{}} (\textbf{\textcolor{violet}{MaNGA}} + \textbf{\textcolor{gray}{HIPASS}} +\\ \textbf{ \textcolor{red}{GAMA} + \textcolor{Orchid}{SDSS} + \textcolor{PineGreen}{REFLEX II}}) \end{tabular}} & \ref{fig:HMF_Integral}                                 & -4.47 $\substack{+0.09 \\ -0.09}$ & 14.39 $\substack{+0.07 \\ -0.09}$ & -1.84 $\substack{+0.01 \\ -0.02}$ & 0.90 $\substack{+0.01 \\ -0.01}$ \\
\midrule
\ \ \ \ \ \  (\textbf{\textcolor{violet}{MaNGA} + \textbf{\textcolor{gray}{HIPASS}}})                                                                                        & -                                   & -3.09 $\substack{+0.03 \\ -0.02}$ & 12.71 $\substack{+0.04 \\ -0.03}$  & -1.58 $\substack{+0.008 \\ -0.01}$ & 1.30 $\substack{+0.01 \\ -0.02}$ \\
\midrule
\multicolumn{1}{l}{\begin{tabular}[c]{@{}c@{}} (\textbf{\textcolor{violet}{MaNGA}}  + \textbf{ \textcolor{red}{GAMA}} \\ + \textbf{ \textcolor{Orchid}{SDSS} + \textcolor{PineGreen}{REFLEX II} })  \end{tabular}} & - & -4.41 $\substack{+0.05 \\ -0.04}$ & 14.33 $\substack{+0.02 \\ -0.03}$ & -1.83 $\substack{+0.01 \\ -0.01}$ & 0.887 $\substack{+0.009 \\ -0.009}$ \\

\midrule
\midrule
\end{tabular*}
\end{table*}

\section{Conclusions}

Using stellar and gas kinematics from the MaNGA galaxy survey, we provide a computation of the CVF in the nearby Universe between $100-350\ \rm{km\ s^{-1}}$ (Fig. \ref{fig:CVF}), for a sample of galaxies with varied morphologies, representative of the nearby-Universe galaxy population with stellar masses above $M_{\star}=10^{9.2}\ \rm{M_{\odot}}$. After careful consideration of observational errors and Eddington bias, we find that the CVF is well described by an MRP function (Fig. \ref{fig:CVF_morph_SF}) and constrain its shape and amplitude. Assuming that circular velocities trace the virial equilibrium of dark matter halos (Fig. \ref{fig:HMF}), we use these metrics to compute the HMF. We combine our HMF measurements with those of \cite{li_halo_2019} and \cite{driver_empirical_2022}, and constrain the functional form of the HMF across 5 orders of magnitude in $M_{200}$ (Fig. \ref{fig:HMF_Integral}). The CVF and HMF derived in this work provide a comprehensive test for $\Lambda$CDM predictions. Our main results are as follows:

\begin{itemize}
    \item \textbf{CVF current state and comparison with literature:} The MaNGA CVF computed in this work (Fig. \ref{fig:CVF}) agrees reasonably well with the computation of \cite{bekeraite_califa_2016} within the range covered by the latter, and extends by an additional $\sim 70\ \rm{km\ s^{-1}}$ to lower velocities, and by $\sim 50\ \rm{km\ s^{-1}}$ to higher velocities. We obtain significantly larger number densities above $\sim 150\ \rm{km\ s^{-1}}$ compared to estimates from \ion{H}{I} surveys (HIPASS and ALFALFA) due to the morphological variety of our sample which includes quenched early-types.

    \item \textbf{CVF modelling and contribution from different morphological and SF classes:} 
    The CVF computed in this work is well-described by an MRP function (Fig. \ref{fig:CVF_morph_SF}). We constrain the shape and amplitude of the CVF and find that LTs dominate the number density of galaxies in the nearby Universe for circular velocities below $250\ \rm{km\ s^{-1}}$, above which ETs are dominant. At $v_{\rm{circ}} \leqslant 200\ \rm{km\ s^{-1}}$ SF galaxies are dominating the nearby-Universe galaxy population. Above this  threshold, SF and NSF galaxies have comparable number densities. 

    \item \textbf{CVF contribution from different dynamical classes:} Dynamically cold discs dominate the number density of galaxies in the nearby Universe for circular velocities in the range $170-315\ \rm{km\ s^{-1}}$ (Fig. \ref{fig:CVF_stellar_gas}). Above $315\ \rm{km\ s^{-1}}$, number densities of cold discs and intermediate systems are comparable. Slow rotators make up a small fraction of the number density of galaxies across our range of probed $v_{\rm{circ}}$, from 1 per cent at $180\ \rm{km\ s^{-1}}$ to 11 per cent at $300\ \rm{km\ s^{-1}}$.

    \item \textbf{Halo mass function from IFS kinematics:} We calculate halo masses for our MaNGA sample using SMHM scaling relations. The resulting HMF for a sample statistically complete above $M_{\star}=10^{9.2}\rm{M_{\odot}}$ (Fig. \ref{fig:HMF}) covers 1 order of magnitude in halo mass ($10^{11.7}-10^{12.7}\rm{M_{\odot}}$) for which HMF measurements were not previously available (or were not complete). The HMF computed in this work agrees remarkably well with the $\Lambda$CDM prediction of \cite{murray_thehalomod_2021}, and with the extrapolation of the \cite{driver_empirical_2022} HMF for group and cluster halos below $10^{12.7} \rm{M_{\odot}}$. SF and NSF galaxies contribute almost equally to the mass density ($\rho_{\rm{M}}$) and the fractional matter density ($\Omega_{\rm{M}}$) in the nearby Universe, with SF galaxies being more numerous below $M_{200}=10^{12.3}\rm{M_{\odot}}$ (Fig. \ref{fig:HMF_split_classes}). Spiral galaxies are contributing 67 per cent to the mass density in the nearby Universe, while lenticulars and ellipticals account for the rest. These findings are representative of the nearby-Universe galaxy population with $M_{\star}\geqslant 10^{9.2}\rm{M_{\odot}}$. When analysing the HMF for our stellar kinematic sample (statistically complete for $M_{\star} \geqslant 10^{10.2}\ \mathrm{M_{\odot}}$, bottom of  Fig. \ref{fig:HMF_split_classes}), we find that kinematically cold discs and inetrmediate systems contribute 97 per cent of the matter density, while the rest corresponds to slow rotators.

    \item  \textbf{Halo mass function constrained across 5 orders of magnitude in $M_{200}$:}
    By combining the HMF data derived in this work with that of \cite{li_halo_2019} and \cite{driver_empirical_2022}, we constrain the functional form of the HMF between $10^{10.5} - 10^{15.5}\rm{M_{\odot}}$ (Fig. \ref{fig:HMF_Integral}). The HMF of galaxies is well modelled by an MRP function, and its integral from zero mass to infinity results in a matter density parameter $\Omega_{\rm{M}}=0.306\substack{+0.018 \\ -0.019}$, consistent with the estimate of \cite{aghanim_planck_2020} without imposing any priors. The joint HMF derived in this work agrees with the $\Lambda$CDM prediction of \cite{murray_thehalomod_2021} within uncertainties. Halos with masses in the range $10^{10.5}\rm{M_{\odot}} - 10^{15.5}\rm{M_{\odot}}$ enclose 72$\substack{+5 \\ -6}$ per cent ($\Omega_{\rm{M,10.5-15.5}} = 0.227 \pm 0.018$) of the matter density in the nearby Universe. The contribution to the matter density from single-occupation halos probed by MaNGA and HIPASS is 31$\substack{+5 \\ -6}$ per cent, the rest being accounted for by group and cluster halos.

\end{itemize}

Our results provide a benchmark computation of the CVF from 100 $\rm{km\ s^{-1}}$ to 350 $\rm{km\ s^{-1}}$, and place strong constraints on the HMF across 5 orders of magnitude in $M_{200}$. These two functions are key statistical probes of the galaxy population in the present day Universe and represent fundamental tests of the  $\Lambda$CDM galaxy evolution framework. While a detailed comparison of these results with outputs of state-of-the art cosmological simulations of galaxy evoution is beyond our scope, we found that the HMF computed in this work agrees well with the $\Lambda$CDM-based computation of \cite{murray_thehalomod_2021}. We placed constraints on the contribution of different galaxy morphological, dynamical and star forming subcategories to the number and mass density of galaxies in the nearby Universe. 

Expanding the CVF to lower circular velocities for morphologically diverse samples will be possible with upcoming large-scale IFS surveys such as HECTOR (\citealt{bryant_hector_2016}). 
Although robustly constraining  the mass density parameter $\Omega_{\rm{M}}$ with the current HMF data is not possible given the required extrapolation (equivalent to 28 per cent of $\Omega_{\rm{M,Planck18}}$), our results show promise for potential progress in this regard from expanding the HMF to lower $M_{200}$. Upcoming surveys such as the 4MOST Wide Area VISTA Extragalactic Survey (WAVES, \citealt{driver_4most_2019}) will result in more robust HMF measurements in the group/cluster regime and facilities such as the Square Kilometer Array (\citealt{dewdney_square_2009}) will allow for an expansion of the HMF from \ion{H}{I} kinematics down to low, uncharted halo masses.

\label{sec:conclusions}

\section*{Acknowledgements}

We thank the anonymous reviewer for a constructive report which improved the quality of this work. \textbf{AR} acknowledges that this research was carried out while the author was in receipt of a Scholarship for International Research Fees (SIRF) and an International Living Allowance Scholarship (Ad Hoc Postgraduate Scholarship) at The University of Western Australia. This research was conducted by the Australian Research Council Centre of Excellence for All Sky Astrophysics in 3 Dimensions (ASTRO 3D), through project number CE170100013.
\textbf{LC} acknowledges support from the Australian Research Council Discovery Project funding scheme (DP210100337). \textbf{DO} is a recipient of an Australian Research Council Future Fellowship (FT190100083) funded by the Australian Government. \textbf{KG} acknowledge support from the Australian Research Council Laureate Fellowship FL180100060. This work has made use of the OzSTAR national computing facility at the Swinburne University of Technology. The OzSTAR program receives funding in part from the astronomy National Collaborative Research Infrastructure Strategy (NCRIS) allocation provided by the Australian government, and from the Victorian Higher Education State Investment Fund (VHESIF) provided by the Victorian Government.

This research has made use of the following Python packages: \textsc{Seaborn} \citep{waskom_seaborn_2021}, \textsc{Matplotlib} \citep{hunter_matplotlib_2007}, \textsc{SciPy} \citep{virtanen_scipy_2020}, \textsc{NumPy} \citep{harris_array_2020}, \textsc{Astropy} \citep{robitaille_astropy_2013}, \textsc{Pingouin} \citep{vallat_pingouin_2018}. We have also used the following R packages: \texttt{dftools} \citep{obreschkow_eddingtons_2018} and \texttt{HYPER-FIT} \citep{robotham_hyper-fit_2015}. 

\section*{Data Availability}

The MaNGA value-added products used in this paper are available at \url{https://www.sdss4.org/dr17/manga/}. Stellar and gas rotational velocities and $v/\sigma$ values in / within 1.3$R_{\rm{e}}$ are published in \cite{ristea_tullyfisher_2024} (Supplementary information). Two catalogues, containing circular velocities and halo masses derived in this work, and the data of the joint HMF in Fig. \ref{fig:HMF_Integral}, are outlined in Appendix \ref{sec:Appendix_catalogue}. Further data can be provided upon request to the author.

 \section*{Author contribution statement}

This project was devised by \textbf{AR}, \textbf{LC}, \textbf{AFM} and \textbf{BG}. \textbf{AR} performed the analysis and drafted the paper.  \textbf{DO} and \textbf{KG} discussed the results and commented on the manuscript.



\bibliographystyle{mnras}
\bibliography{References_Paper_3} 





\section*{Supporting information}
Additional supporting information can be found in the online version of this article.

\textbf{Two data catalogues are provided with this work:} 
\begin{itemize}
    \item \textit{HMF\_combined\_data.fits/.csv}  
    \item \textit{MaNGA\_DR17\_Kinematic\_vcirc\_M200\_Ristea24.fits/.csv}
\end{itemize}
Refer to Appendix \ref{sec:Appendix_catalogue} for details on each catalogue. 

We also present a \textit{.gif} version of Fig. \ref{fig:sigmoid_3d} (viewed in different projections), \textit{Fig\_4\_3D.gif}.

\appendix

\section{Completeness of samples used in this work}
\label{sec:Appendix_completeness}

\begin{figure}
	\centering
	\includegraphics[width=\columnwidth]{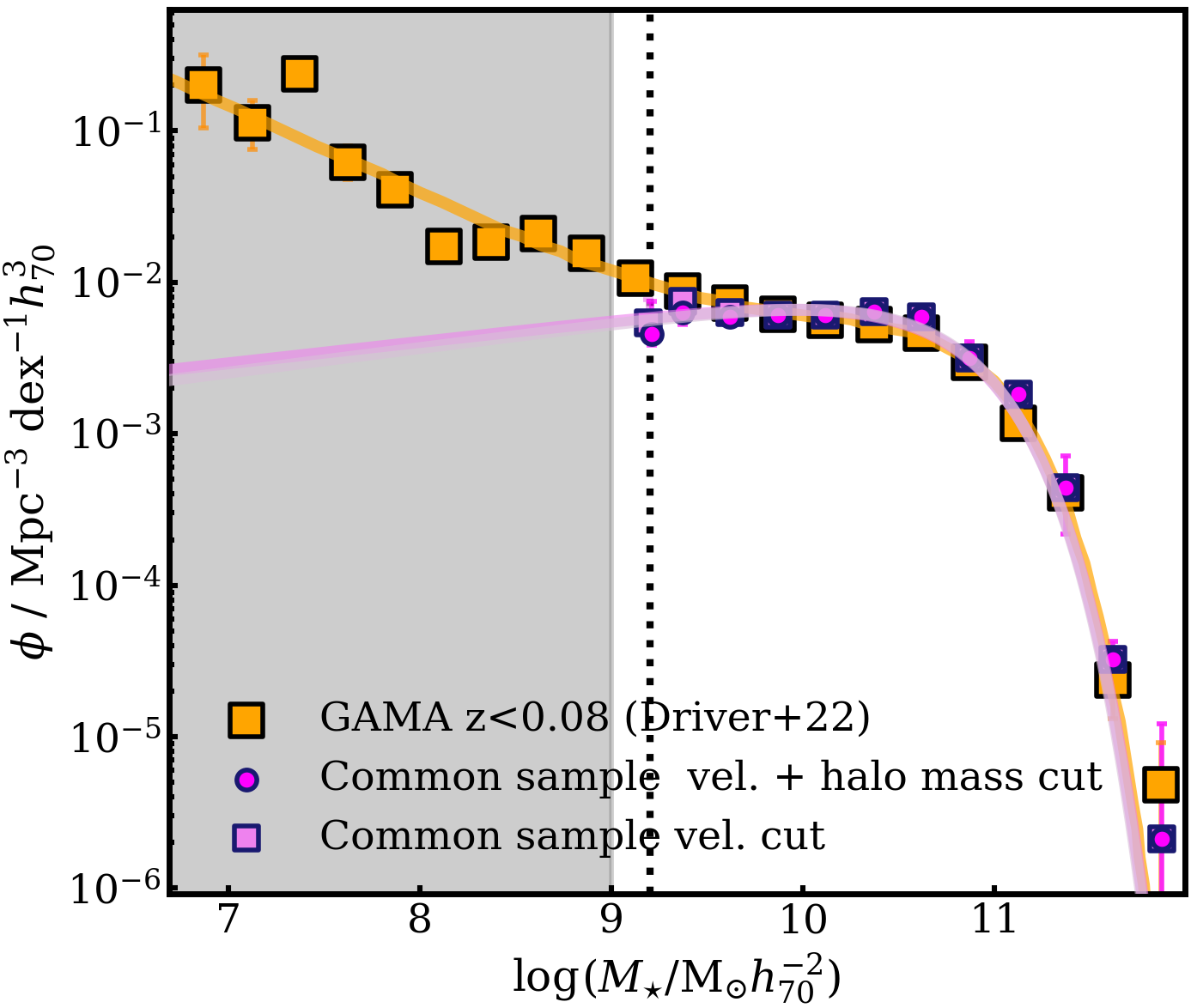}
    \caption{Similar to Fig. \ref{fig:MF } (right), now showing the SMF of our MaNGA common sample after the cut in circular velocity $\log_{10}(v_{\rm{circ}}/\rm{km\ s^{-1}})\geqslant 2.011$ (pink squares), and after the further cut in halo mass $\log_{10}(M_{200}/\rm{M_{\odot}})\geqslant 11.35$ (pink circles). The pink curves show Schechter fits to the two SMFs. After the implementation of the two cuts, the agreement with the GAMA SMF is only maintained for $\log_{10}(M_{\star}/\rm{M_{\odot}}) \geqslant 9.2$ (black dotted line). As such, the results related to the circular velocity and halo mass functions in Fig. \ref{fig:CVF}, \ref{fig:CVF_morph_SF}, \ref{fig:HMF}, \ref{fig:HMF_split_classes} and  \ref{fig:HMF_Integral} should be interpreted as reflecting a galaxy sample statistically complete above $10^{9.2} \rm{M_{\odot}}$.       } 
    \label{fig:MF_Additional}
\end{figure}

In Table \ref{tab:samples_completeness} we present a description of the various galaxy samples used in this work, in terms of number of galaxies and completeness. The initial common kinematic sample (purple contours in Fig. \ref{fig:sample_plot}) is representative of the nearby Universe galaxy population above $10^{9} \ \rm{M_{\odot}}$. In Section \ref{sec: Circular velocity function} we apply a cut in circular velocity and only retain galaxies with $\log_{10}(v_{\rm{circ}}/ \rm{km\ s^{-1}})\geqslant 2.011$, below which our sample is not statistically complete. The resulting sample matches the GAMA SMF above $10^{9.2}\ \rm{M_{\odot}}$. This comparison is illustrated in Fig. \ref{fig:MF_Additional}. In Section \ref{sec: SMHM rel} we apply a further cut and only keep galaxies with $\log_{10}(M_{200}/\rm{M_{\odot}})\geqslant10^{11.35} \ \rm{M_{\odot}}$. This exclusion does not further reduce the completeness of the final sample used for HMF computation (complete above $10^{9.2}\  \rm{M_{\odot}}$, Fig. \ref{fig:MF_Additional}). After every cut applied to the common kinematic sample, effective volumes are re-calculated as described in Section \ref{sec:volume weights}. All samples used in this work are complete up to $M_{\star} = 10^{11.9}\ \rm{M_{\odot}}$. 

\begin{table*}

\centering
\caption{A description of the galaxy samples used in this work. The "\textit{Complete above $\log_{10}(M_{\star}/\rm{M_{\odot}})$}" column refers to the lower limit in stellar mass above which the SMF of the sample reproduces that of the GAMA survey (\citealt{driver_galaxy_2022}). The cut in $\log_{10}(v_{\rm{circ}})$ was applied to the common kinematic sample to exclude the range of velocities within which our CVF was declining due to a lack of completeness (Fig. \ref{fig:CVF}). The further cut in $\log_{10}(M_{\rm{200}})$ was applied to exclude the region in the SMHM relation within which the running median through our data shows a much lower gradient than expected from theoretical modeling and previous observations (Fig. \ref{fig:stellar_mass_halo_mass}). All samples are complete up to $M_{\star}=10^{11.9}\rm{M_{\odot}}$. The full \textbf{\textcolor{violet}{common}}, \textbf{\textcolor{red}{stellar}} and \textbf{\textcolor{blue}{gas}} samples were cut in $\log_{10}(M_{\star}/\rm{M_{\odot}})$ to only include galaxies with masses above the value in the third column.  }
\setlength\tabcolsep{1.9pt}
\renewcommand{\arraystretch}{1.2}
\label{tab:samples_completeness}
\begin{tabular*}{\linewidth}{@{\extracolsep{\fill}}c|c|c|c}

Sample                  & Number of galaxies \ \ \ \ \ \ \ \ \ \   & Complete above $\log_{10}(M_{\star}/\rm{M_{\odot}}) =$ \ \ \ \ \ \ \ \ & Appears in figures \\
\hline
\hline
\textbf{\textcolor{violet}{Common}} (full sample)   &  3527 \ \ \ \ \ \ \ \ \ \    &        9.0 \ \ \ \ \ \ \ \ \ \               &   Fig. \ref{fig:sample_plot}, \ref{fig:TF_circ_vel}, \ref{fig:MF }, \ref{fig:CVF}        \\

\textbf{\textcolor{violet}{Common}} with   $\log_{10}(v_{\rm{circ}}/\rm{km\ s^{-1}}) \geqslant 2.011$            &  \centering 3456 \ \ \ \ \ \ \ \ \  \ &    9.2   \ \ \ \ \ \ \ \ \ \                  &   Fig. \ref{fig:CVF_morph_SF}, \ref{fig:stellar_mass_halo_mass}                \\

\begin{tabular}[c]{@{}c@{}}  \textbf{\textcolor{violet}{Common}} with  $\log_{10}(v_{\rm{circ}}/\rm{km\ s^{-1}}) \geqslant 2.011$ \\ \textbf{and}  $\log_{10}(M_{\rm{200}}/\rm{M_{\odot}}) \geqslant 11.35$
\end{tabular}

&  3378   \ \ \ \ \ \ \ \ \ \ &     9.2      \ \ \ \ \ \ \ \ \ \              & Fig. \ref{fig:HMF}, \ref{fig:HMF_split_classes}, \ref{fig:HMF_Integral} \\                 
\hline

\textbf{\textcolor{red}{Stellar}}   &  1850    \ \ \ \ \ \ \ \ \ \ &        10.1  \ \ \ \ \ \ \ \ \ \               &   Fig. \ref{fig:sample_plot}, \ref{fig:sigmoid_2d} (bottom), \ref{fig:sigmoid_3d} (left), \ref{fig:MF }, \ref{fig:CVF_stellar_gas}        \\

\textbf{\textcolor{red}{Stellar}} with $\log_{10}(M_{200}/\mathrm{M_{\odot}}) \geqslant 11.8$   &  1673    \ \ \ \ \ \ \ \ \ \ &        10.2  \ \ \ \ \ \ \ \ \ \               &   Fig. \ref{fig:HMF_split_classes}      \\

\textbf{\textcolor{blue}{Gas}}    & 2834   \ \ \ \ \ \ \ \ \ \  &        9.0   \ \ \ \ \ \ \ \ \ \              &   Fig. \ref{fig:sample_plot}, \ref{fig:sigmoid_2d} (top), \ref{fig:sigmoid_3d} (right), \ref{fig:MF }, \ref{fig:CVF_stellar_gas}        \\
\hline

\end{tabular*}
\end{table*}

\section{Systematic uncertainties in $M_{200}$}
\label{sec:Appendix_errors}

As detailed in \cite{mo_formation_1998}, the mass of a dark matter halo $M_{200}$ can be computed from the circular velocity $v_{\mathrm{circ}} \equiv v_{200}$, the Hubble constant $H(z)$ and the gravitational constant $G$ as:

\begin{equation}
    M_{200}(v_{\mathrm{circ}}) = \frac{v_{\mathrm{circ}}^3  }{10GH(z)}.
    \label{eq:m200_2}
\end{equation}

The halo mass can also be calculated from the virial radius $R_{200}$ as: 

\begin{equation}
    M_{200}(R_{200}) = \frac{100H^{2}(z) R^{3}_{200}  }{G}.
    \label{eq:m200_3}
\end{equation}

\begin{figure}
	\centering
	\includegraphics[width=\columnwidth]{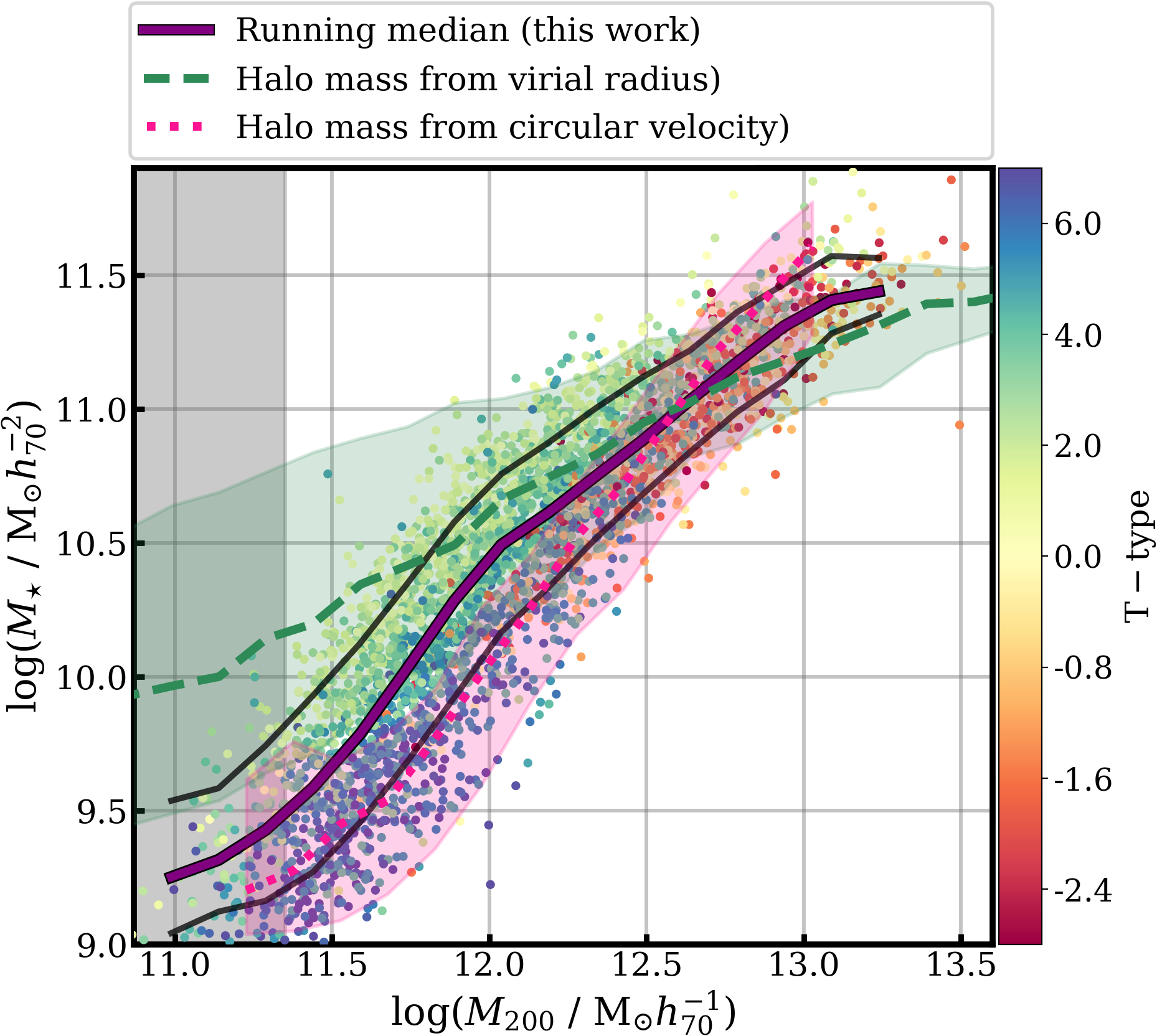}
    \caption{Similar to Fig. \ref{fig:stellar_mass_halo_mass}, displaying the running medians of the SMHM relations computed using equations \ref{eq:m200_2} (pink) and \ref{eq:m200_3} (green), and the corresponding $16^{\mathrm{th}}$ and $84^{\mathrm{th}}$ percentiles (shaded areas).       } 
    \label{fig:Appendix_SMHM}
\end{figure}

The approach followed in this work (equation \ref{eq:m200}) is chosen due to its reliability in reproducing literature SMHM relations (Fig. \ref{fig:stellar_mass_halo_mass}). The $M_{200}$ computations from equations \ref{eq:m200_2} and \ref{eq:m200_3} are only equivalent if $v_{\mathrm{circ}} = 10 R_{200}H(z)$. However, given that $R_{200}$ and $v_{\mathrm{circ}}$ are estimated independently, this relation will not exactly hold. As such, the assumption of a specific analytical relation for computing $M_{200}$ is a source of systematic error. Differences between the various methods of computing $M_{200}$ are highlighted in Fig. \ref{fig:Appendix_SMHM}, on the SMHM plane The running medians are marginally consistent within the $16^{\mathrm{th}}$ - $84^{\mathrm{th}}$ percentile confidence intervals, with differences between the methods becoming larger at lower masses.  

Equation \ref{eq:m200} results in halo mass values which are equal to the geometrical mean of those given by equations \ref{eq:m200_2} and \ref{eq:m200_3}, if the former carries twice the weight of the latter (and is thus always bounded by the two estimates). We consider the upper and lower systematic uncertainty in our $M_{200}$ estimates as:

\begin{equation}
 \alpha_{\mathrm{sys}} =
\begin{cases}
       \mathrm{max}\big[M_{200}(\mathrm{\ref{eq:m200_2}}), M_{200}(\mathrm{\ref{eq:m200_3}})  \big] - M_{200}(\mathrm{\ref{eq:m200}}) \ \ \ \  (\mathrm{upper});  \\
        M_{200}(\mathrm{\ref{eq:m200}})  - \mathrm{min}\big[M_{200}(\mathrm{\ref{eq:m200_2}}), M_{200}(\mathrm{\ref{eq:m200_3}})  \big] \ \ \ \ \ (\mathrm{lower}).
    \end{cases}\,
\end{equation}

To assess the effect of systematic uncertainties on the HMF derived in this work, we re-compute the HMFs in Fig. \ref{fig:HMF} and Fig. \ref{fig:HMF_Integral} for $M_{200} + \alpha_{\mathrm{sys,upper}}$ and $M_{200} - \alpha_{\mathrm{sys,lower}}$. The resulting best-fitting MRP functions and density parameter values are shown in Table \ref{tab:systematic_effect}.

For the MaNGA HMF (Fig. \ref{fig:HMF}), the change in the best-fitting parameters when systematic uncertainties are taken into account is within 8 per cent (for both upper and lower $\alpha_{\mathrm{sys}}$) in all cases.
The same modification in best-fitting HMF parameters for the joint HMF in Fig. \ref{fig:HMF_Integral} is found to be within 2 and 1 per cent (for upper and lower $\alpha_{\mathrm{sys}}$, respectively) in all cases. The $\Omega_{\mathrm{M,10.5-15.5}}$ parameter (within the probed range of halo masses) changes by 11 and 12 per cent when the upper and lower systematic errors are considered, while the same change for the $\Omega_{\mathrm{M}}$ parameter (integrated down to zero halo mass) is 8 per cent (for both  the upper and lower uncertainties). Changes in the $\rho_{\rm{M,sample}}$ and $\Omega_{\rm{M,sample}}$ values in Table \ref{tab:rho_m_split} are found to be within 9 per cent, in all cases.

\begin{table}
\centering
\caption{ The best-fitting MRP parameters of the joint HMF derived in this work, and resulting matter density parameters, after the upper ($\alpha_{\mathrm{sys,upper}}$) and lower ($\alpha_{\mathrm{sys,lower}}$) systematic uncertainties are taken into account. The values from Fig. \ref{fig:HMF_Integral} are also shown, under the $M_{200}$(\ref{eq:m200}) column for comparison. }
\label{tab:systematic_effect}
\setlength{\tabcolsep}{1.64pt}

\begin{tabular}{@{\extracolsep{\fill}}c|c|c|c}

\multicolumn{1}{c}{} & \multicolumn{3}{c}{Halo mass} \\
\midrule 
 Parameter & $M_{200}$(\ref{eq:m200}) & \ \ \ \ \ \ \ \ \ \ \ \  +$\alpha_{\mathrm{sys,upper}}$ & -$\alpha_{\mathrm{sys,lower}}$  \\
\midrule
\midrule
\multicolumn{4}{c}{\textbf{\textcolor{violet}{MaNGA}} \textbf{HMF}  (Fig. \ref{fig:HMF}) } \\
\midrule
\midrule
$\log_{10}(\phi_{*}/\mathrm{Mpc^{-3} }h_{\mathrm{67}}^{3})$           &  -2.97              & -2.76  & -3.21        \\

 $\log_{10}(M_{*}/\mathrm{M_{\odot} }h_{67}^{-1})$          &   12.69               & 11.90 & 12.65            \\
$\alpha$              &    -1.54            &   -1.44 & -1.67         \\

$\beta$        &    1.26            & 1.15 & 1.33           \\
\midrule
\midrule
\multicolumn{4}{c}{ (\textbf{\textcolor{violet}{MaNGA}} + \textbf{\textcolor{gray}{HIPASS}} + \textbf{ \textcolor{red}{GAMA}} + \textbf{\textcolor{Orchid}{SDSS}} + \textbf{\textcolor{PineGreen}{REFLEX II}} ) \textbf{HMF} (Fig. \ref{fig:HMF_Integral}) } \\
\midrule
\midrule
$\log_{10}(\phi_{*}/\mathrm{Mpc^{-3} }h_{\mathrm{67}}^{3})$           &  -4.47              & -4.46  & -4.50       \\

 $\log_{10}(M_{*}/\mathrm{M_{\odot} }h_{67}^{-1})$          &   14.39               & 14.43 & 14.33            \\
$\alpha$              &   -1.84             &   -1.83 & -1.85         \\

$\beta$        &    0.90            & \ 0.91 & 0.88           \\
\midrule
$\Omega_{\mathrm{M,10.5-15.5}}$                  &  0.227              & \  0.253 & \  0.199        \\
$\Omega_{\mathrm{M}}$ (integrated to 0 mass)                 &  0.306              & \   0.331  & \   0.280       \\
\midrule
\midrule
\end{tabular}
\end{table}

\section{Kinematic and halo mass catalogue ($v_{\rm{circ}}$ and $M_{200}$) and combined HMF data}
\label{sec:Appendix_catalogue}

We present a compilation of the joint data sets used to construct the HMF in Fig. \ref{fig:HMF_Integral}. This data is presented in the form of \textit{.fits} and \textit{.csv} files (\textit{HMF\_combined\_data.fits/.csv}, available online only). Table \ref{tab:functions_data} below presents the formatting of these files, showing the first data point of each of the surveys used in the joint HMF. The $M_{200}$ values and number densities in these data tables assume a Planck 2018 cosmology (\citealt{aghanim_planck_2020}, as in Fig. \ref{fig:HMF_Integral}).

\begin{table*}
\centering
\caption{Format of the data table \textit{HMF\_combined\_data.fits/.csv}, outlining the data of the joint halo mass function (units are the same as in Fig. \ref{fig:HMF_Integral}).   }
\setlength\tabcolsep{1.9pt}
\renewcommand{\arraystretch}{1.2}
\label{tab:functions_data}
\begin{tabular*}{\linewidth}{@{\extracolsep{\fill}}ccccccc}
Survey   & log\_M200 & err\_log\_M200\_lower & err\_log\_M200\_upper & number\_density & err\_number\_density\_lower & err\_number\_density\_upper \\
\midrule
\midrule
HIPASS   & 10.4556   &                       &                       & 0.038913        & 0.01792                     & 0.02509                     \\

MaNGA    & 11.54590   & 0.11385              & 0.06422             & 0.010595        & 5.641e-4                   & 5.5896e-4                   \\
SDSS     & 12.8740   &                       &                       & 1.1885e-3       & 2.3921e-4                   & 2.3728e-4                   \\
GAMA     & 13.3262   &                       &                       & 1.2469e-3       & 3.1880e-4                   & 3.4076e-4                   \\

REFLEX II & 12..7275  &                       &                       & 7.7179e-4       & 6.8058e-4                   & 7.2447e-4                  
\end{tabular*}
\end{table*}

We also present a catalogue of the circular velocities ($v_{\rm{circ}}$) and halo masses ($M_{\rm{200}}$) of MaNGA DR17 galaxies derived in this work. The catalogue is provided in the form of \textit{.fits} and \textit{.csv} files (\textit{MaNGA\_DR17\_Kinematic\_vcirc\_M200\_Ristea24.fits/.csv}, available online only). These files are formatted in the same way as Table \ref{tab:catalogue}, which presents data for 5 example galaxies. The table includes the following columns:
\begin{itemize}
\item \textbf{plate\_ifu}: Plate-ifu of the observation, the same as in \texttt{drpall$\_$v3$\_$1$\_$1};
  \item \textbf{Manga\_ID}: MaNGA ID of the galaxy, the same as in \texttt{drpall$\_$v3$\_$1$\_$1};
  \item \textbf{RA\_deg}: Right ascension [deg];
  \item \textbf{DEC\_deg}: Declination [deg];
  \item \textbf{log\_Mstar}: $\log_{10}$ of the stellar mass, in $\mathrm{M_{\odot}}h_{70}^{-2}$;
   \item \textbf{log\_SFR}: $\log_{10}$ of the star formation rate in $\mathrm{M_{\odot}\ yr^{-1}}h_{70}^{-2}$;
   \item \textbf{T\_type}: T-type of the galaxy;
   \item \textbf{log\_vcirc}: $\log_{10}$ of the circular velocity (in $\rm{km\ s^{-1}}$) derived in this work (Section \ref{sec:circular velocities});
   \item \textbf{err\_log\_vcirc}: uncertainty in the $\log_{10}$ of the circular velocity (in $\rm{km\ s^{-1}}$) derived in this work;
   \item \textbf{log\_M200}: $\log_{10}$ of the halo mass (in $\mathrm{M_{\odot}}h_{70}^{-1}$) derived in this work (Section \ref{sec:HMF_IFS});
   \item \textbf{err\_log\_M200}: uncertainty in the  $\log_{10}$ of the halo mass (in $\rm{M_{\odot}}h_{70}^{-1}$) derived in this work;
   \item \textbf{z\_spec}: Spectroscopic redshift;
   \item \textbf{z\_min}: Minimum redshift at which a galaxy could have been observed given the MaNGA selection function;
   \item \textbf{z\_max}: Maximum redshift at which a galaxy could have been observed given the MaNGA selection function;
   \item \textbf{Sample}: \textbf{1} if the galaxy is in the full common kinematic sample, \textbf{2} if the galaxy has $\log_{10}(v_{\rm{circ}}/\rm{km\ s^{-1}}) \geqslant 2.011$, \textbf{3} if the galaxy has $\log_{10}(v_{\rm{circ}}/\rm{km\ s^{-1}}) \geqslant 2.011$ \textbf{and} $\log_{10}(M_{200}/\mathrm{M_{\odot}}h_{70}^{-1}) \geqslant 11.35$.
\end{itemize}


\bsp	
\label{lastpage}

\begin{landscape}
\begin{table}

\caption{Format of the data table \textit{MaNGA\_DR17\_kinematic\_vcirc\_M200\_Ristea24.fits/.csv}, outlining the kinematic and halo mass catalogues derived in this paper.}
\begin{tabular}{ccccccccccccccc}
\label{tab:catalogue}
plate\_ifu   & Manga\_ID & RA\_deg      & DEC\_deg    & log\_Mstar & log\_SFR & T\_type & log\_vcirc & err\_log\_vcirc & log\_M200 & err\_log\_M200 & z\_spec     & z\_min & z\_max  & Sample \\
\midrule
\midrule
9192-9102 & 1-37606 & 46.23777 & 0.21681 & 11.066 & -0.194      & 1.09187   & 2.41077       & 0.00855            & 12.34088     & 0.01727           & 0.02896 & 0.03473  & 0.06296 & 3      \\
      11956-6103       &    1-151429     &  187.17995       &   53.35796     & 10.917      &  -0.469   &   0.65072     &    2.42796        &   0.05388              &   12.2884        &    0.05596            &   0.03658    &  0.03473      &  0.06296     &   3     \\
       9182-3701     &  1-604915        &   120.13796      &   39.84006     &  9.909     &  0.092   &    1.63991    &    2.21863        &   0.0349              &     11.73937      &        0.03801        &  0.02809     &  0.03472      &   0.06296    &    3    \\
        10215-6103    &   1-122139       &  122.32324       &   38.33694     & 10.009           & 0.202   &   2.0149  &    1.98997        &     0.21264            &   11.25623        &    0.21331            &   0.04072    & 0.09074       &  0.10052     &   1     \\
        11017-1901    &     1-319663     &    202.52767     &  50.51109      & 9.806        & -1.897   &    0.10898    &  2.11187          &     0.05241            &    11.32821       &     0.05454           &  0.02568     &   0.03473     &    0.06296   &     2 \\  
\end{tabular}
\end{table}
\end{landscape}

\end{document}